\documentclass{article}

\usepackage[cmex10]{amsmath}
\usepackage{amsfonts}
\usepackage{amsmath}
\usepackage{amssymb}
\usepackage{setspace}
\usepackage{color}
\usepackage{subfigure}
\usepackage{subfigmat}
\usepackage{graphicx}
\usepackage{booktabs}
\usepackage{array,hhline}
\usepackage{hyperref}
\usepackage{float}

\graphicspath{{figs/}}

\title{Multi-vehicle Path Following using Modified Trajectory Shaping Guidance}

\author{Ishmaal T. Erekson$^{1}$ Rajnikant Sharma$^{2}$ Ashwini Ratnoo$^{3}$ and Ryan Gerdes $^{4}$
\thanks{*This work was not supported by any organization}
\thanks{$^{1}$Ishmaal Erekson  is with Sandia National Laboratories, NM
        {\tt\small erekson89@gmail.com}}%
\thanks{$^{2}$Rajnikant Sharma is with Department of Aerospace Engineering, University of Cincinnati,
        Cinicnnati, OH 45221, USA
        {\tt\small rajnikant.sharma@uc.edu}}%
        \thanks{$^{3}$Ashwini Ratnoo is with Department of Aerospace Engineering, Indijintan Institute of Science,
        Bangalore, India
        {\tt\small ratnoo@aero.iisc.ernet.in}}%
        \thanks{$^{4}$ Ryan Gerdes is with Department of Electrical and Computer Engineering, Virginia Tech,
        Virginia, USA
        {\tt\small rgerdes@vt.edu}}%
}

\begin{document}

\maketitle
\thispagestyle{empty}
\pagestyle{empty}

\begin{abstract}

In this paper, we formulate a  virtual target-based path following guidance law aimed towards multi-vehicle path following problem. The guidance law is well suited  to precisely follow circular paths while minting desired distance between two adjacent vehicles where path information is only available to the lead vehicle.  We analytically show lateral and longitudnal stability and convergence on the path.  This is also validated through simulation and experimental results. 

\end{abstract}

\section{INTRODUCTION}

With fast evolving levels of autonomy, autonomous vehicles seek accurate and easily implementable path following guidance methods. Tracking a virtual target on the desired path provides a promising path following guidance framework. A nonlinear guidance logic for path following was proposed by Park et al.\cite{Park2004,park2007performance} wherein the vehicle uses an instantaneous circular maneuver to pursue a virtual target moving at a constant look-ahead distance on the path. The resulting linear analysis of the system shows an asymptotic convergence to straight line and circular paths. It was observed that a small fixed look-ahead distance very small noise could lead to very high maneuvers and a higher fixed look ahead leads to high With a pure pursuit based guidance law, Medagoda and Gibbens~ \cite{Medagoda2010} used a new virtual target motion logic where the virtual target speed varies with its relative distance to the vehicle. This adaptive approach overcomes the problems arising due to a fixed look-ahead distance. 

Missile guidance theory comprises a rich collection of empirical and optimal guidance laws.  Using linearized kinematics, Ryoo et al.~\cite{Ryoo2005}  derived the optimal trajectory shaping missile guidance law against stationary targets. The shaping effect is characterized by the terminal angle of approach at the target. The trajectory shaping guidance command is a function of vehicle heading, line-of-sight angle and the desired terminal approach angle. In a recent work, Ratnoo et al.~\cite{Ratnoo2015} used the trajectory shaping guidance law for unmanned vehicle path following considering a virtual target on the path. The work shows that even in highly offset initial conditions, trajectory shaping guidance outperforms pure pursuit in terms of path-following errors and also has a faster rate of convergence as compared nonlinear guidance logic. However, as acknowledged in~\cite{Ratnoo2015} , the trajectory shaping convergence property for curved paths holds only for small look ahead distances relative to the path radius. As the look-ahead distance to path radius ratio increases the vehicle converges to a larger offset froPolytechnic Institute and State Universitym the path radius. In a single vehicle path following scenario the look ahead distance to the virtual target is a design/tuning parameter.  For multiple vehicle path following, or platooning, the look ahead distance is an input for maintaining a desired inter-vehicle separation, which could be high due to safety and stability reasons. Also, certain road or weather conditions may also require large look-ahead distances. In a platoon, due to the nature of each vehicle following the one in front of itself, any path following offset amplifies as it propagates through to the end of the platoon. This motivates a need to address the limitation of the trajectory shaping algorithm in accurately converging to path for any desired look ahead distance. The existing platooning work~\cite{Rajamani2000, Fritz1999} require reference markerPolytechnic Institute and State Universitys and and marker detection capability on each vehicle for path following. In this paper, we use trajectory shaping guidance concept to platooning where only the lead vehicle require path knowledge and other vehicles can follow the path by just following the vehicle in front of it.  Although the approach in this paper seems similar to the cyclic pursuit concept~\cite{pavone2007decentralized,galloway2009geometry,kim2007cooperative}, both have different outcomes. Cyclic pursuit is a formation control problem with three different formations, namely, rendezvous to a point, evenly spaced circular formation, and evenly spaced logarithmic spirals. However, in addition to precise path convergence, the multiple vehicle path following problem in this paper achieves precise and desired separation on a circular path.

The main contributions of this paper are as follows. First, the convergence properties of the trajectory shaping guidance algorithm are investigated.  Its dependence on $d^*$ and a nonzero offset from the path are shown. Secondly, the trajectory shaping guidance algorithm is modified to follow curved paths at zero offset. Finally, the modified trajectory shaping law is then applied to control a platoon of vehicles by making each vehicle the virtual target of the vehicle behind it.  It is shown both analytically and with the help of experiments on a multi-robot test-bed that the platoon converges to  a constant radius path. 

This rest of the paper is organized in the following manner. Section~\ref{sec:mod} discusses the regular and modified trajectory shaping guidance law. Section~\ref{sec:plat} presents platooning using modified trajectory shaping guidance and an convergence analysis of the proposed method.  Section~\ref{sec:plat_res} shows and discusses the simulation and hardware results of the modified trajectory shaping applied to a single vehicle and a platoon. Concluding remarks are presented in Section~\ref{sec: conclusions}.

\section{Trajectory Shaping Guidance}
	\label{sec:mod}
\begin{figure}[!h]
	\centering
	\subfigure[Trajectory Shaping Guidance Geometry 1]{\includegraphics[width = 0.48\textwidth]{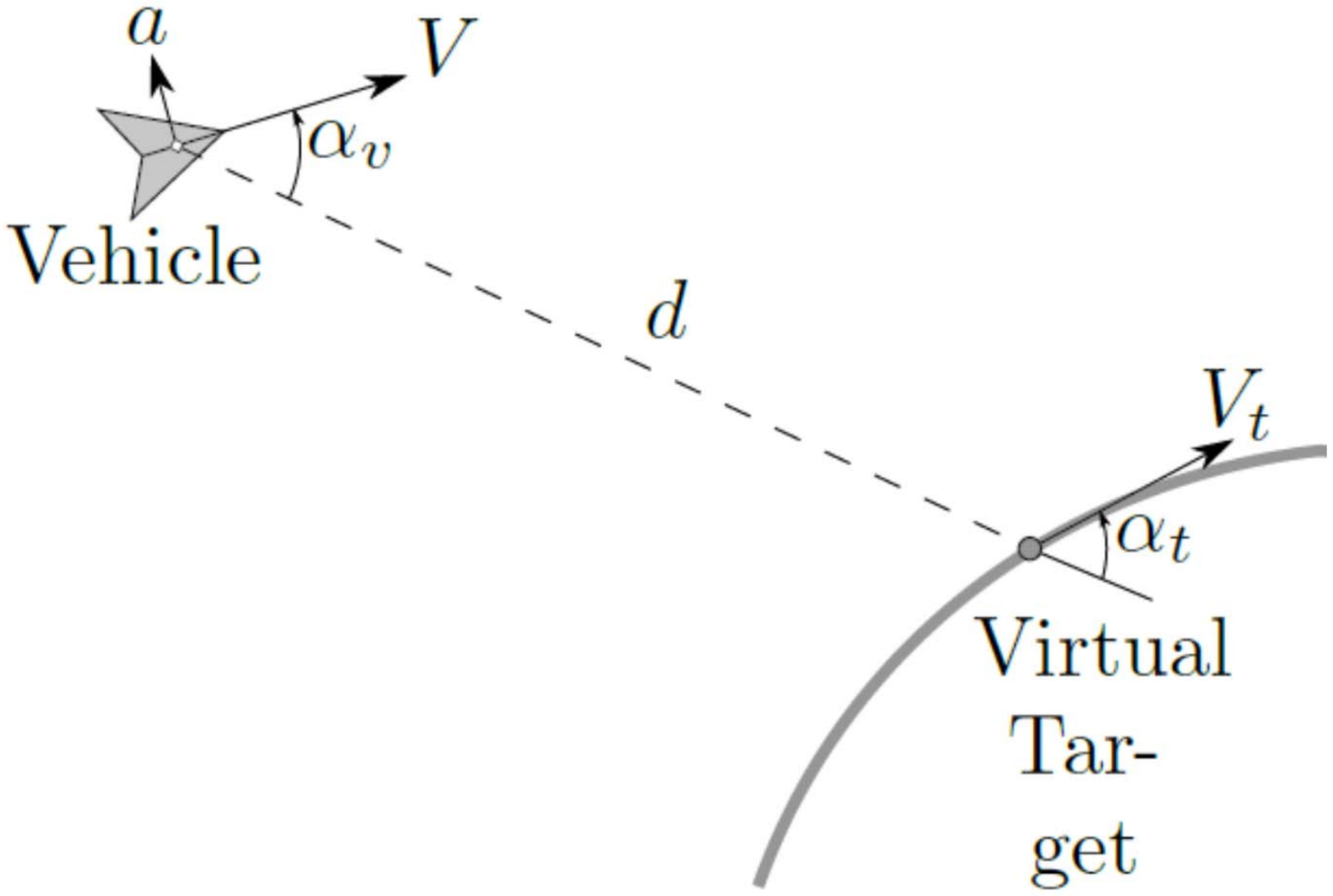}}\label{fig:TS}
	\subfigure[Trajectory Shaping Guidance Geometry 2]{\includegraphics[width = 0.48\textwidth]{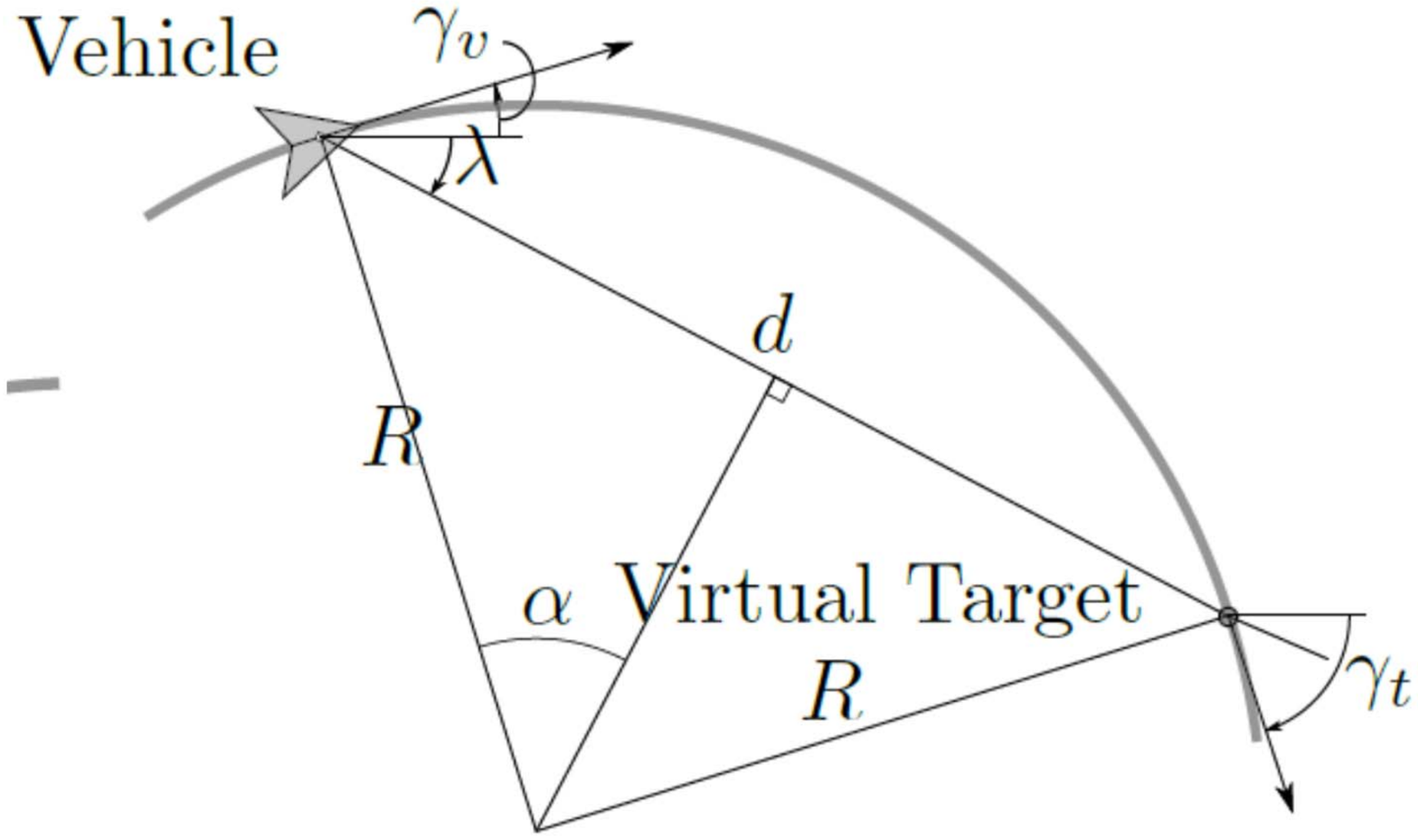}}
		\label{fig:TSsingleVehicle}
	\caption{Trajectory Shaping Guidance Geometry} 
\end{figure}

	Consider a vehicle moving at  velocity $V$ and a virtual target moving on a given path at velocity $V_t$ as shown  in Fig.~\ref{fig:TS}.  Trajectory shaping guidance law governs the vehicle's heading direction so as to achieve a tail-chase approach to the virtual target which is constrained to move along the desired path.  The virtual target moves at a distance $d$ ahead of the vehicle.  The virtual target moves with a relatively lower speed drawing the vehicle closer to the path. As the distance $d$ reduces, the virtual target speed increases and finally matches the vehicle's speed at the desired  distance $d^*$.  The states of the system are $x=[d,~\alpha_t,~\alpha_v, V]$ where $\alpha_t=\gamma_t-\lambda$ is relative target heading,  $\alpha_v= \lambda - \gamma_v$ is the relative vehicle heading,  here $\lambda$ is line-of-sight angle, $\gamma_v$ and $\gamma_t$ are vehicle and virtual target heading angles, respectively. The input vector is $u=[\frac{1}{R},~V_c]^\top$ where $\frac{1}{R}$ is the curvature of the path and $V_c$ is the commanded velocity of the vehicle following the path. The equations of motion can be written as 
	\begin{align}\label{e: ss1}
		\dot{x}=\left[\begin{array}{c}
		\dot{d}\\ 
		\dot{\alpha}_{t}  \\ 
		\dot{\alpha}_{v}\\
		\dot{V}
		\end{array}\right] =f(x,u)=\left[\begin{array}{c}
		V_t\cos\alpha_{t}-V\cos\alpha_{v}\\
		\frac{V_t}{R}-\frac{( V_t\sin\alpha_{t}-V\sin\alpha_{v})}{d}\\
		\frac{a}{V}-\frac{( V_t\sin\alpha_{t}-V\sin\alpha_{v})}{d}\\
		k_v(V_c-V)
		\end{array}\right],
	\end{align}
	where $k_v$ is velocity controller gain. 		

Trajectory shaping guidance~\cite{Ratnoo2015} will be referred to in this paper as regular trajectory shaping guidance.  It controls the vehicle's lateral acceleration $a$ and the virtual target velocity $V_t$ using the following equations 
\begin{align}\label{e: Regular TS}
a &= \frac{V^2}{d} \left(4(\lambda - \gamma_v) + 2(\lambda - \gamma_t)\right),
\end{align}
\begin{align}\label{e: Vt}
V_t &= V \frac{d^*}{d},
\end{align}
Note that, the speeds become identical as the vehicle reaches the desired relative distance $d^*$ following the virtual target.

	   It was shown that the regular trajectory shaping guidance  achieves convergence to a straight line and near convergence to a circular path. However, this is under the assumption of small relative heading angles $\gamma_v$ and $\gamma_t$ obtained for $d << 2R$.   It can be shown that the vehicle will actually converge to an offset of the path radius when $d^* << 2R$ is not met. Fig.~\ref{fig:TSsingleVehicle} shows that the desired equilibrium point for the vehicle to be on the circular path at a desired distance $d^*$ is the following.
	\begin{align} \label{e: eq point 1}
	x^{eq}=\left[\begin{array}{c}
                  d^*\\ 
	             \sin^{-1}\frac{d^*}{2R}\\ 
                 -\sin^{-1}\frac{d^*}{2R}\\
                 V_c
	\end{array}\right],~
		u^{eq}=\left[\begin{array}{c}
		\frac{1}{R}\\ 
		V_c
		\end{array}\right]
		\end{align}
		
	However, it can be shown that for regular trajectory shaping~\eqref{e: Regular TS} and~\eqref{e: ss1},  $\dot{x}=f(x^{eq},u^{eq})=0$ for straight line $R=\infty$, but for circular paths $\dot{x}=f(x^{eq},u^{eq})\neq 0$.  This means that unless the path is a straight line, the equilibrium point of the regular method is not the desired equilibrium point and thus the vehicle will not converge to the path at the desired distance.  
	
\subsection{Modified Trajectory Shaping Guidance}
	In order to obtain accurate path convergence, the lateral acceleration command of the regular trajectory shaping guidance in \eqref{e: Regular TS} is modified as follows.
	\begin{equation}\label{eq:mtraj}
		a= \frac{V^2}{d} \left(4\sin(\lambda - \gamma_v) + 2\sin(\lambda - \gamma_t)\right).
	\end{equation}
	As can be seen, the only modification from~\eqref{e: Regular TS} is to take the sine of the two angles of interest.  Throughout this paper the modified method will be referred to as modified trajectory shaping or the sine method. It can be easily verified that at  $x^{eq}$ and $u^{eq}$ from~\eqref{e: eq point 1} for modified trajectory shaping law $\dot{x}=f(x^{eq},u^{eq})=0$ for both circular and straight line trajectories. 
This shows that the desired equilibrium point is an actual equilibrium point for the sine method which will result in zero offset. 
 
  It will be also shown in the next section that the system around the desired equilibrium point is stable and the vehicle converges on the path with zero offset independent of the $d^*$ value. The modified trajectory shaping will be used for multiple vehicle path following/platooning in the next section. 
\section{Platooning Using Modified Trajectory Shaping}
	\label{sec:plat}
\begin{figure}[!h]
		\centering
		\includegraphics[scale=0.6]{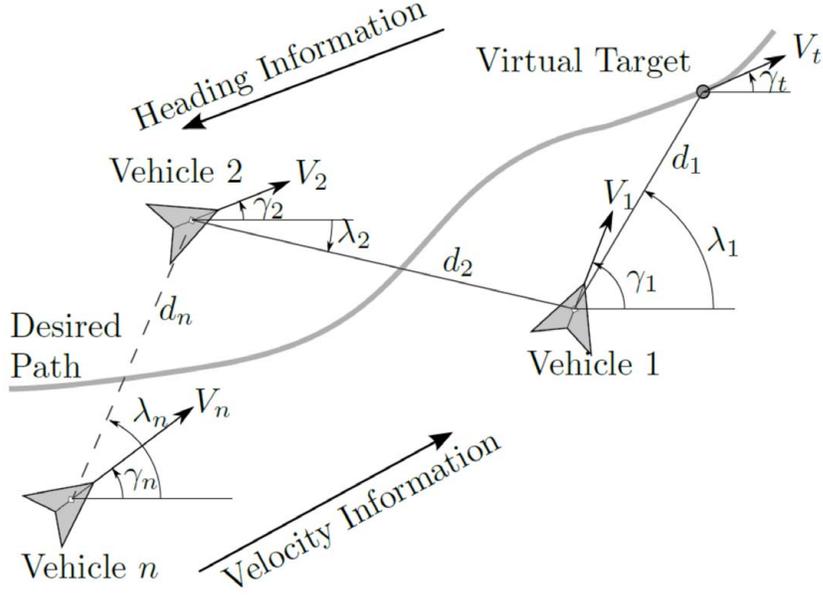}
		\vspace{-7 cm}
		\caption{Platooning Trajectory Shaping}
		\label{fig:veh}
	\end{figure}

		In this section we use the modified/sine trajectory shaping method for platooning. Consider $n$ vehicles and a path to follow as shown in Fig.~\ref{fig:veh}. We assume that the first vehicle has knowledge of the path and it follows a virtual target moving on the path. Each vehicle $i=2,\cdots~n$ follows the vehicle in front of it  ($(i-1)^{th}$ vehicle) as a virtual target.  	
	
	The state vector for a $n$ vehicle platoon is  $x=[x_1,\cdots, x_n]^\top \in \mathbf{R}^{4n}$ where $x_i=[d_i,~\alpha_{t_i},~\alpha_{v_i},~V_i]$. The input vector is $u=[\frac{1}{R},~V_c]$ includes the curvature of the path and the commanded velocity $V_c$ of the last vehicle.  The overall equations of the motion are written as
		\begin{align}\label{e: full ss}
		\dot{x}=\left[\begin{array}{c}
		\dot{x}_1\\ 
	\vdots\\ 
		\dot{x}_n
		\end{array}\right] =f(x,u)=\left[\begin{array}{c}
		f_1(x_i,u)\\ 
		\vdots\\ 
		f_n(x_i,u)
		\end{array}\right]
		\end{align}
	
	State space equation for the $i^{th}$ vehicle can be written as
	\begin{align}\label{e: ss actuated}
	\dot{x}_i=\left[\begin{array}{c}
	\dot{d}_i \\ 
	\dot{\alpha}_{t_i}  \\ 
	\dot{\alpha}_{v_{i}}\\
	\dot{V}_i
	\end{array}\right] =\left[\begin{array}{c}
	V_{i-1}\cos\alpha_{t_i}-V_{i}\cos\alpha_{v_i}\\
	\frac{a_{i-1}}{V_{i-1}}-\frac{( V_{i-1}\sin\alpha_{t_i}-V_{i}\sin\alpha_{v_i})}{d_i}\\
	\frac{a_{i}}{V_{i}}-\frac{( V_{i-1}\sin\alpha_{t_i}-V_{i}\sin\alpha_{v_i})}{d_i}\\
	k_v(V_{cmd_i}-V_i)
	\end{array}\right], 
	\end{align}
	where $V_0=V_t=V_1\frac{d^*}{d_1}$, $a_0=a_t=\frac{V_t^2}{R}$, $V_{cmd_i}=V_{i+1}\frac{d^*}{d_i}$, $V_{cmd_n}=V_c$, and
	 \begin{align*}
	 a_i=\frac{V_i^2}{d_i}(-4\sin\alpha_{v_i}-2\sin\alpha_{t_i}).
	 \end{align*}  The lateral acceleration  command is the lateral controller to minimize path error and the velocity controller is the longitudinal controller for maintaining desired separation between two vehicles. These control equations show (also shown in Fig.~\ref{fig:veh}) that the information needed for the lateral acceleration commands are passed backward through the platoon.  The information needed for velocity control are passed forward through the platoon. This asymmetric information flow provides some benefits to the system when dealing with disturbances.  For example, if a vehicle receives a lateral disturbance from the path, the vehicles in front of that vehicle won't be affected laterally, they will remain on the path.  If a vehicle receives a longitudinal disturbance, the vehicles behind it won't be affected longitudinally.  This is revisited in the simulation and hardware results of Section \ref{sec:plat_res}.
	 
	
\subsection{Platooning Stability Analysis}

For stability analysis we use linearization around the equilibrium point on the circular path. It can be easily verified that at equilibrium on the circular path
\begin{align*}
d_i^{eq}=d^*,~\alpha_{t_i}^{eq}=\sin^{-1}\frac{d^*}{2R},~
\alpha_{v_i}^{eq}=-\alpha_{t_i}^{eq},~
V_t^{eq}=V_i^{eq}=V_c.
\end{align*}
Linearizing \eqref{e: full ss} at the equilibrium point gives	$\dot{\bar x}=A\bar x$ where $\bar x=x-x^{eq}$ is the deviation from the equilibrium point and $A=\frac{\partial f}{\partial x}(x^{eq},~u^{eq})$ is an almost block diagonal (staircase) $4n\times 4n $ matrix
\begin{align}
A=\left[\begin{array}{cccccc}
A_{tt} & A_{tv} &  &  & & \\
A_{vt} & A_{vv} & A_{tv}   &  & & \\
& A_{vt} & A_{vv}  & A_{tv} & & \\
&  & A_{vt} & A_{vv}  & A_{tv} &  \\
&  &  & \ddots  & \ddots &  \ddots
\end{array}\right]
\end{align}
where 
\begin{align*} 
A_{tt}=\left[\begin{array}{cccc}
-\frac{V_c\alpha}{d^*} & -\frac{V_c d^*}{2R} & -\frac{V_c d^*}{2R} & 0\\
\frac{V_c}{2Rd^*}  & -\frac{V_c\alpha}{d^*} & \frac{V_c\alpha}{d^*}& 0\\
\frac{V_c}{2Rd^*}  & -\frac{3V_c\alpha}{d^*} & -\frac{3V_c\alpha}{d^*} & 0\\
0 & 0 & 0 & -k_v 
\end{array}\right],\\
A_{vv}=\left[\begin{array}{cccc}
0 & -\frac{V_c d^*}{2R} & -\frac{V_c d^*}{2R} & -\alpha\\
\frac{V_c}{Rd^*}  & -\frac{V_c\alpha}{d^*} & \frac{V_c\alpha}{d^*}&  -\frac{1}{2R}\\
0  & -\frac{3V_c\alpha}{d^*} & -\frac{3V_c\alpha}{d^*} & \frac{1}{2R}\\
0 & 0 & 0 & -k_v 
\end{array}\right]\\
A_{tv}=\left[\begin{array}{cccc}
0 & 0 & 0 & 0\\
0  & 0 & 0 &  0\\
0  & 0 & 0 &0\\
-\frac{V_ck_v}{d^*} & 0 & 0 & k_v 
\end{array}\right], \\
A_{vt}=\left[\begin{array}{cccc}
0 & 0 & 0 & \alpha\\
-\frac{V_c}{Rd^*}  & -\frac{2V_c\alpha}{d^*}  & -\frac{4V_c\alpha}{d^*} &  \frac{1}{2R}\\
0  & 0 & 0 & -\frac{1}{2R}\\
0 & 0 & 0 & 0
\end{array}\right],
\end{align*}
where $\alpha=\sqrt{1-\left(\frac{d^*}{2R}\right)^2}$.  Since $\frac{d^*}{2R} \leq 1$, $0 < \alpha \leq 1$ and is real. Due to the asymmetric information flow $A_{tv}\neq A_{vt}$. There are a total of $4n$ eigenvalues of $A$ and the system will exponentially converge to the equilibrium point on the path when disturbed if all the $4n$ eigenvalues have negative real parts. After analysis and algebraic manipulations the following $4n$ eigenvalues are computed

\begin{align}\label{e: eigenvalues}
\nonumber \lambda_1&=-k_v,~
\lambda_2=-\frac{V_c\alpha}{d^*},\\
\nonumber \lambda_{3},~\cdots ,~\lambda_{n+2} &=-\frac{2V_c\alpha}{d^*}+ j V_c\sqrt{2\left(\frac{\alpha^2}{d^{*2}}+\frac{1}{4R^2}\right)},\\
 \lambda_{n+3},,~\cdots,~\lambda_{2n+2} &=-\frac{2V_c\alpha}{d^*}- j V_c\sqrt{2\left(\frac{\alpha^2}{d^{*2}}+\frac{1}{4R^2}\right)},\\
\nonumber \lambda_{2n+3},~\cdots ,~\lambda_{3n+1}&=-(1-\beta)k_v\\
\nonumber \lambda_{3n+2},~\cdots,~\lambda_{4n}&=-\beta k_v,
\end{align}
where $0<\beta<1$ is a function of $k_v$, for $k_v=0.5, ~ \beta=0.95$.  It can be seen that the above $4n$ eigenvalues only depend on $k_v, \alpha, \beta,  V_c, d^*$ and are independent of the number vehicles $n$. Since real parts of all the eigenvalues are negative, we can say that the platoon is exponentially stable  in the neighborhood of the equilibrium point on the path. This means that if the platoon is disturbed the vehicles will converge back to the path.

\section{Results}
\label{sec:plat_res}
This section discusses the results obtained through simulations and hardware experiments using the modified trajectory shaping guidance law to platooning control.  Simulation results will first be analyzed to show path convergence of every vehicle in the platoon.  Similar tests were performed on Pololu m3pi robots and also show path convergence of all vehicles.
\subsection{Simulation Results}
\begin{figure}[!h]
	\centering
	\subfigure[Regular Method Velocity Errors]{\includegraphics[width = 0.48\textwidth]{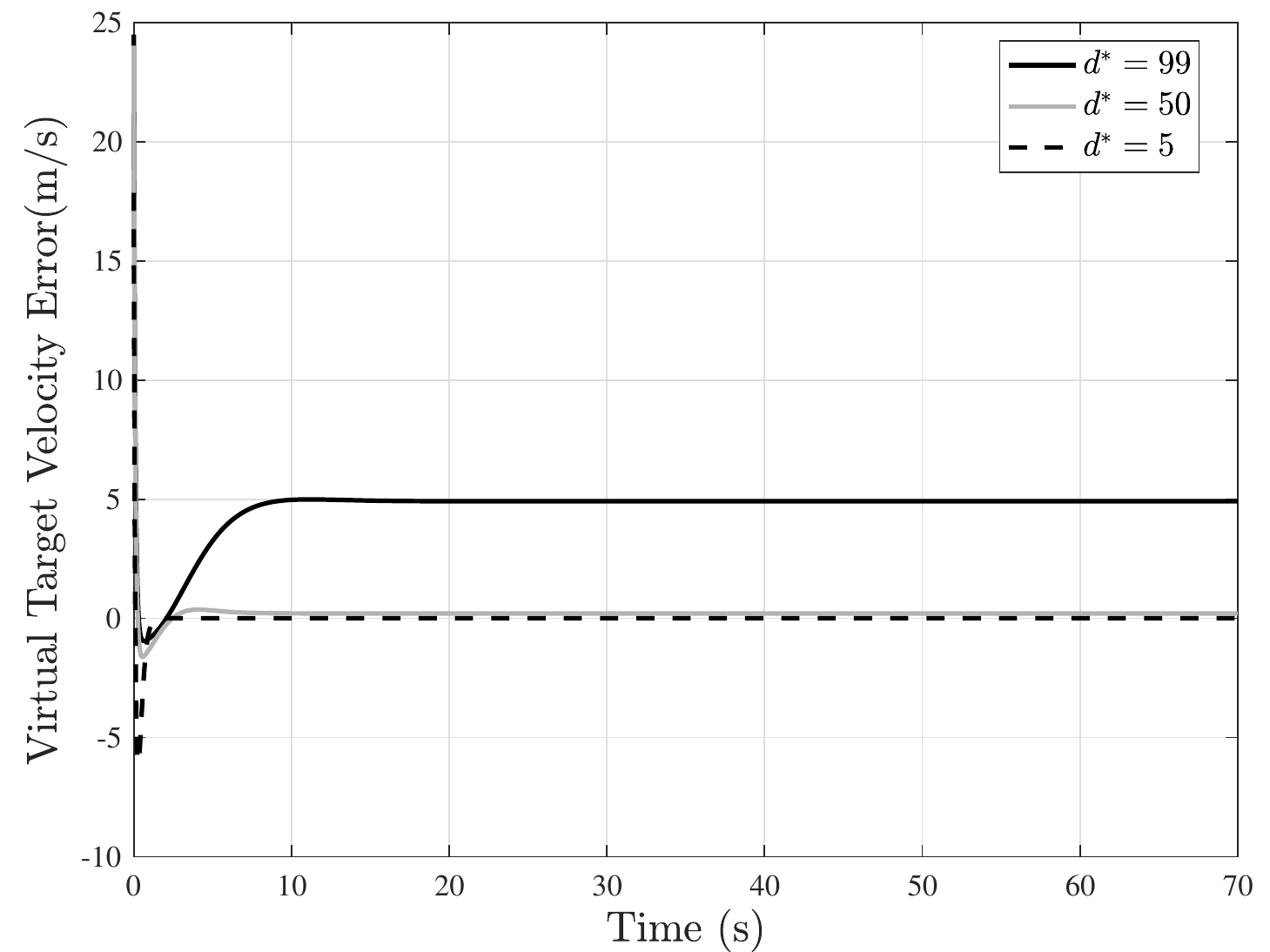}}
	\subfigure[Sine Method Velocity Errors]{\includegraphics[width = 0.48\textwidth]{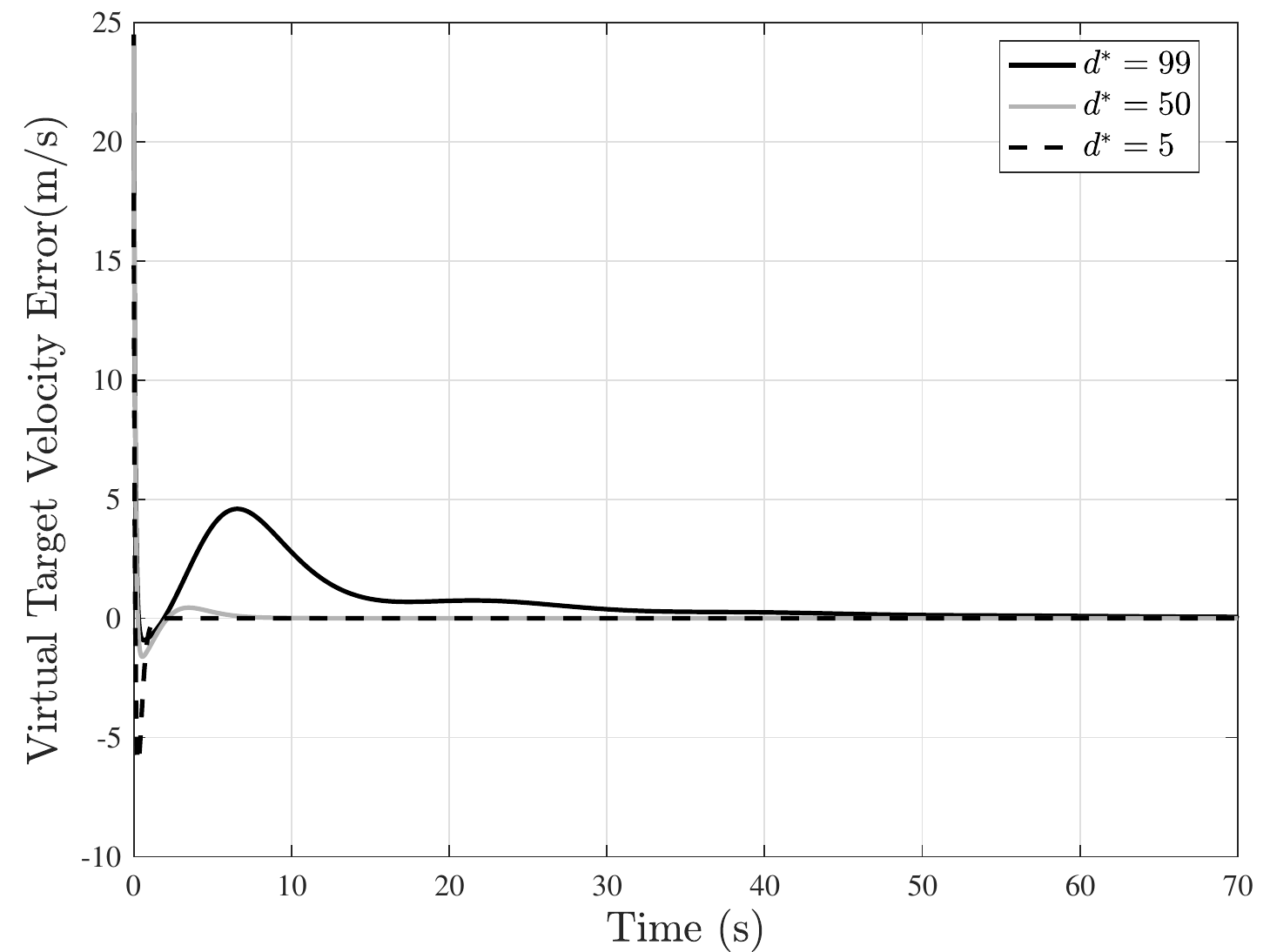}}
    \subfigure[Regular Method Distance Errors Errors]{\includegraphics[width = 0.48\textwidth]{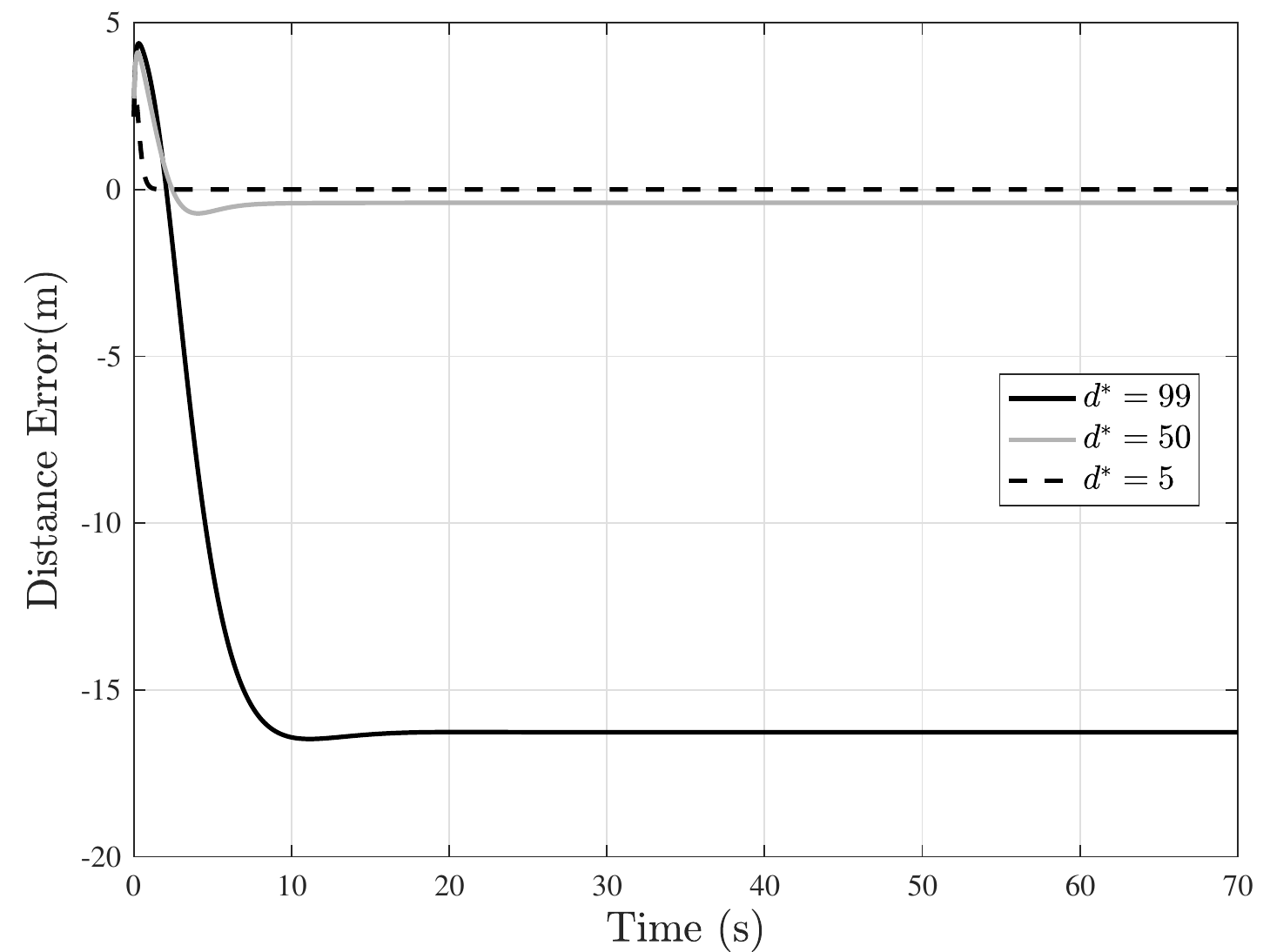}}
    \subfigure[Sine Method Distance Errors]{\includegraphics[width = 0.48\textwidth]{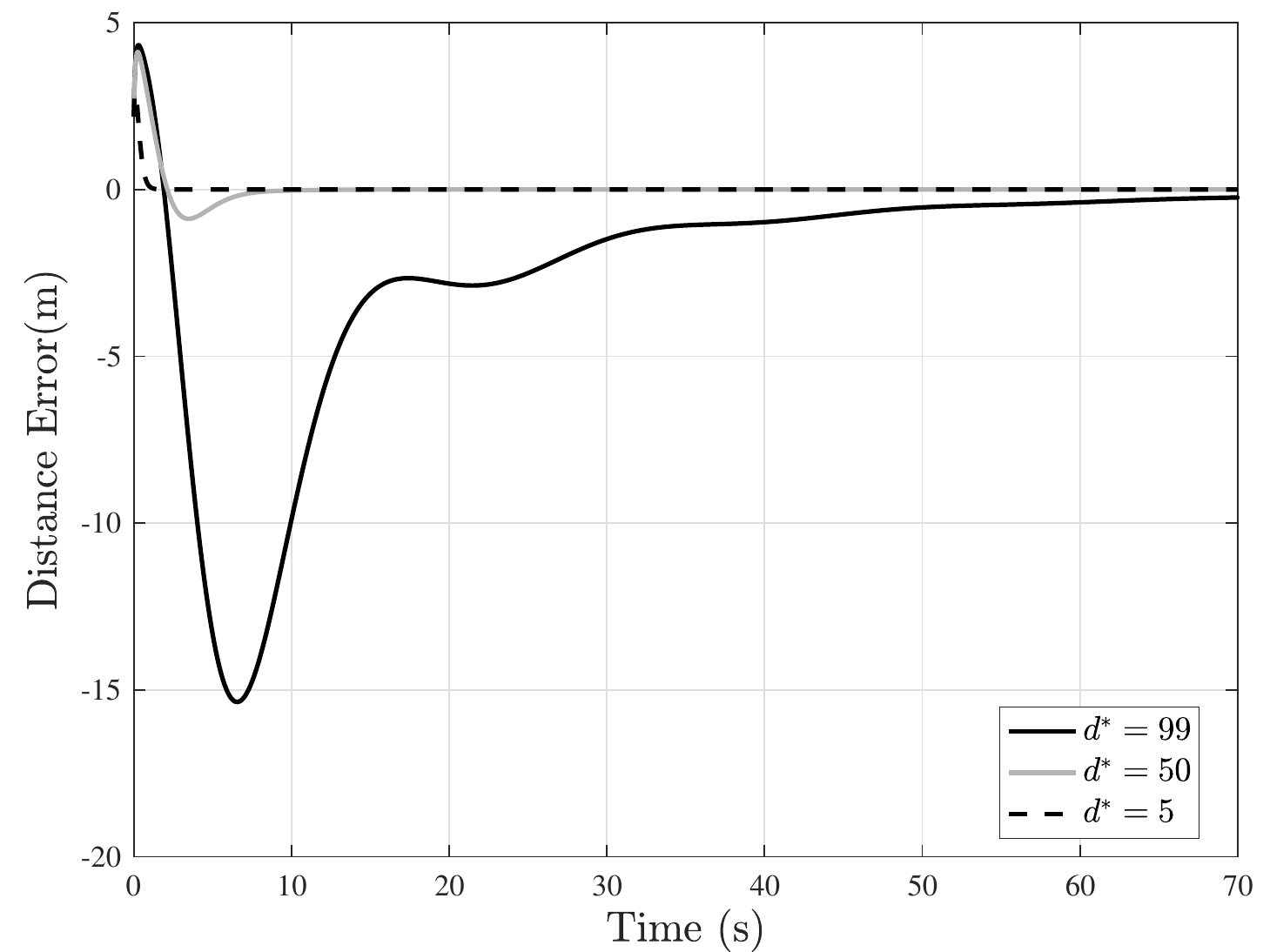}}
    \subfigure[Regular Method Path Errors]{\includegraphics[width = 0.48\textwidth]{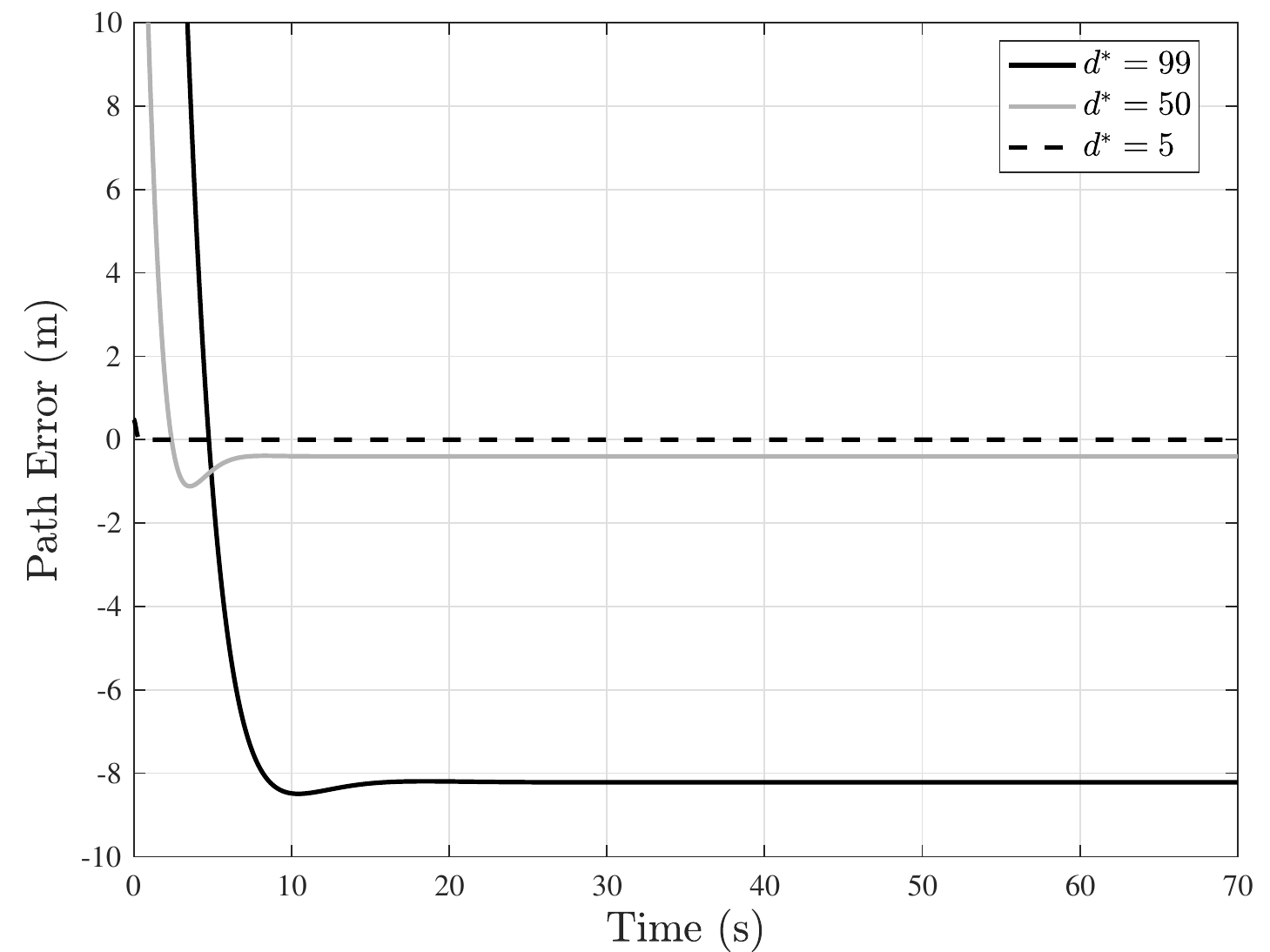}}
    \subfigure[Sine Method Path Errors]{\includegraphics[width = 0.48\textwidth]{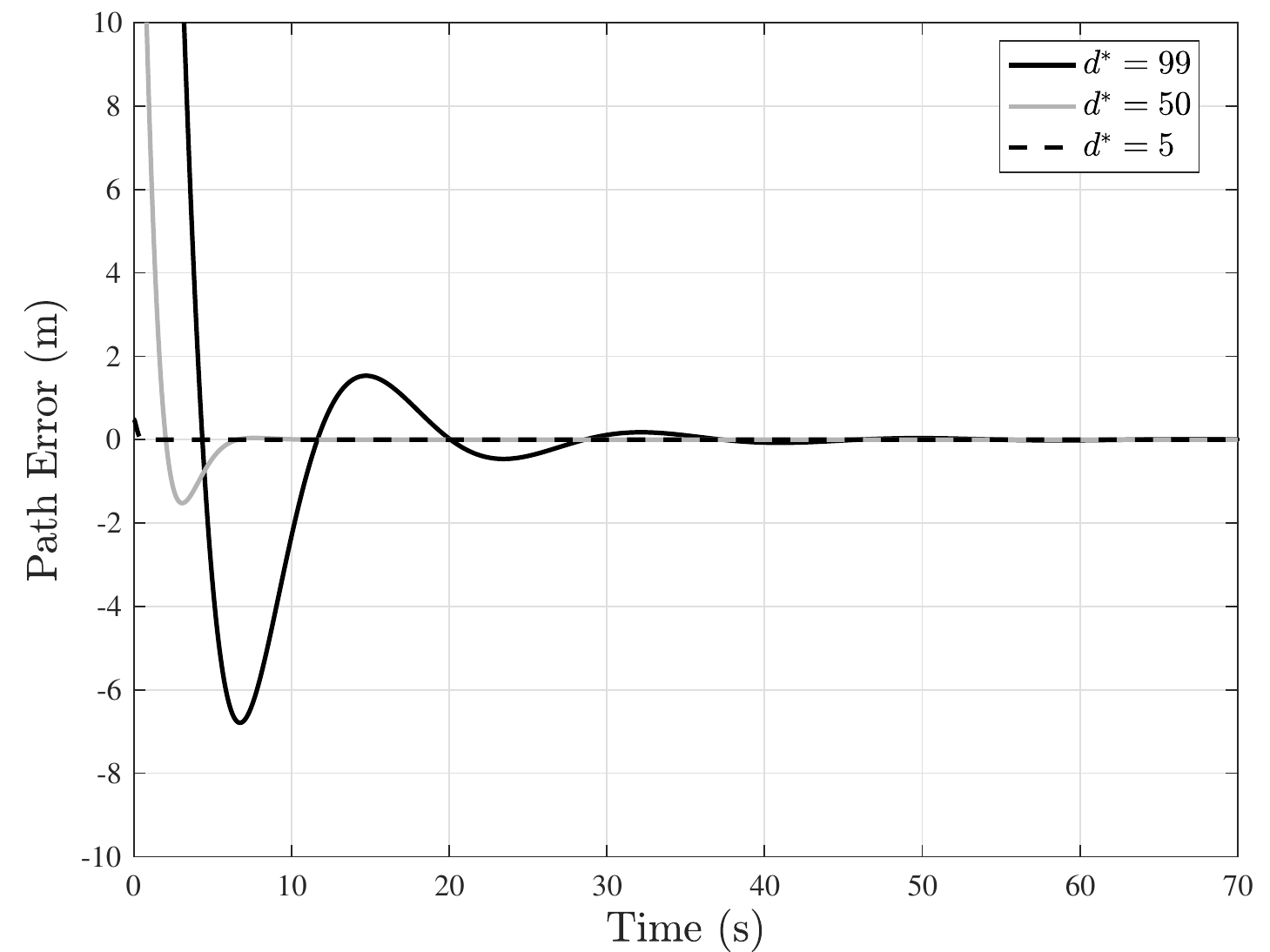}}
	\caption{Single Vehicle Simulation Results}
	\label{fig:comp_sim}
\end{figure}

The simulations were created using the same framework as used for the simulations in~\cite{Erekson2016T}.  However, the simulations in this paper more closely relate to highway conditions.  Each vehicle is treated as the target for the vehicle behind it.  The simulations run for this section were performed with the following parameters: $R = 50~\textrm{m}$, $d^* = 75~\textrm{m}$ (unless otherwise specified), and $V_n = 25~\textrm{m}/s$.  However to first test the effectiveness of the sine method, several scenarios were simulated for a single vehicle as shown in Fig.~\ref{fig:comp_sim}.

\begin{figure}[!h]
	\centering
	\subfigure[Regular Method Velocities]{\includegraphics[width = 0.48\textwidth]{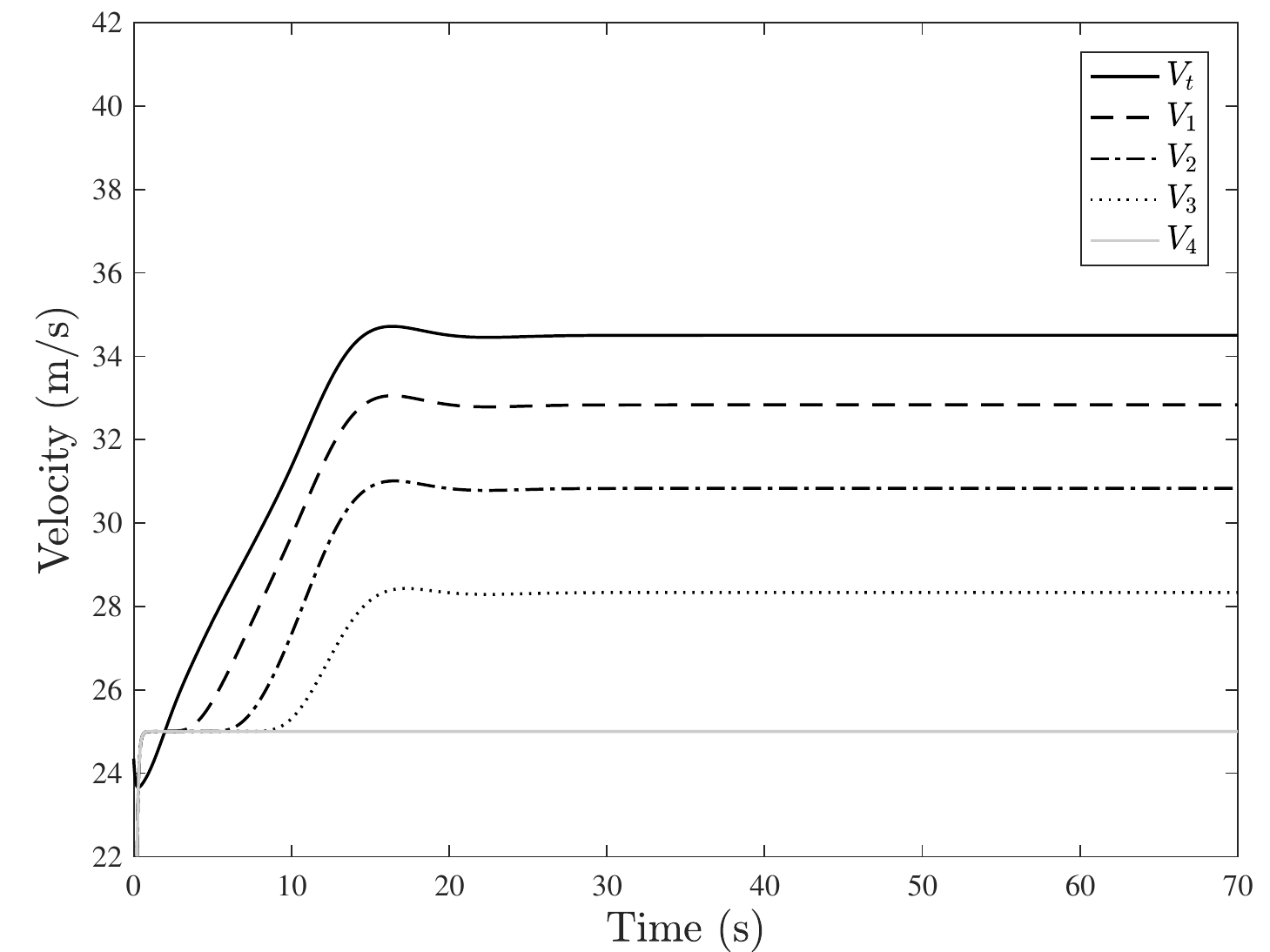}}
	\subfigure[Sine Method Velocities]{\includegraphics[width = 0.48\textwidth]{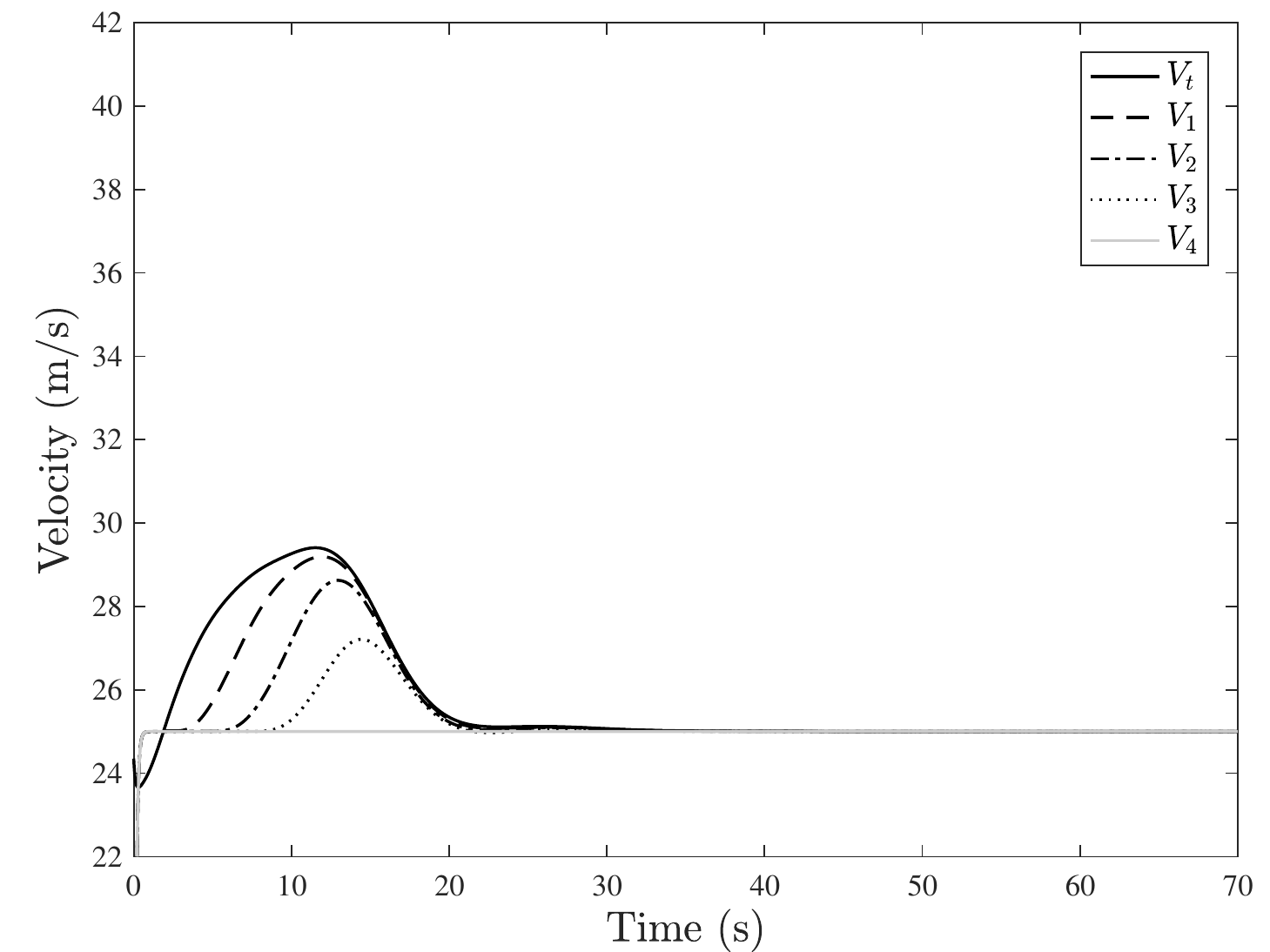}}
	\subfigure[Regular Method Distances]{\includegraphics[width = 0.48\textwidth]{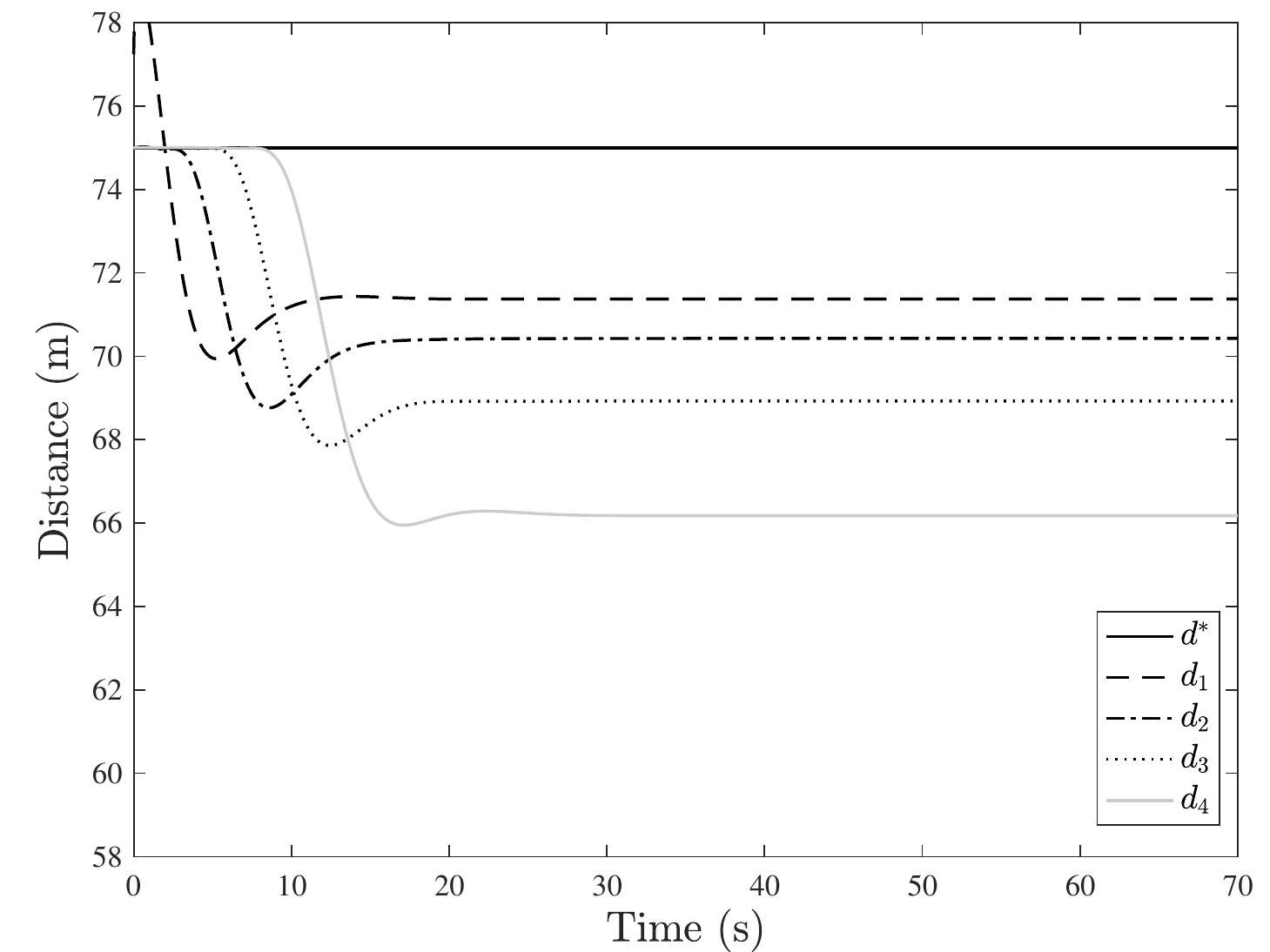}}
	\subfigure[Sine Method Distances]{\includegraphics[width = 0.48\textwidth]{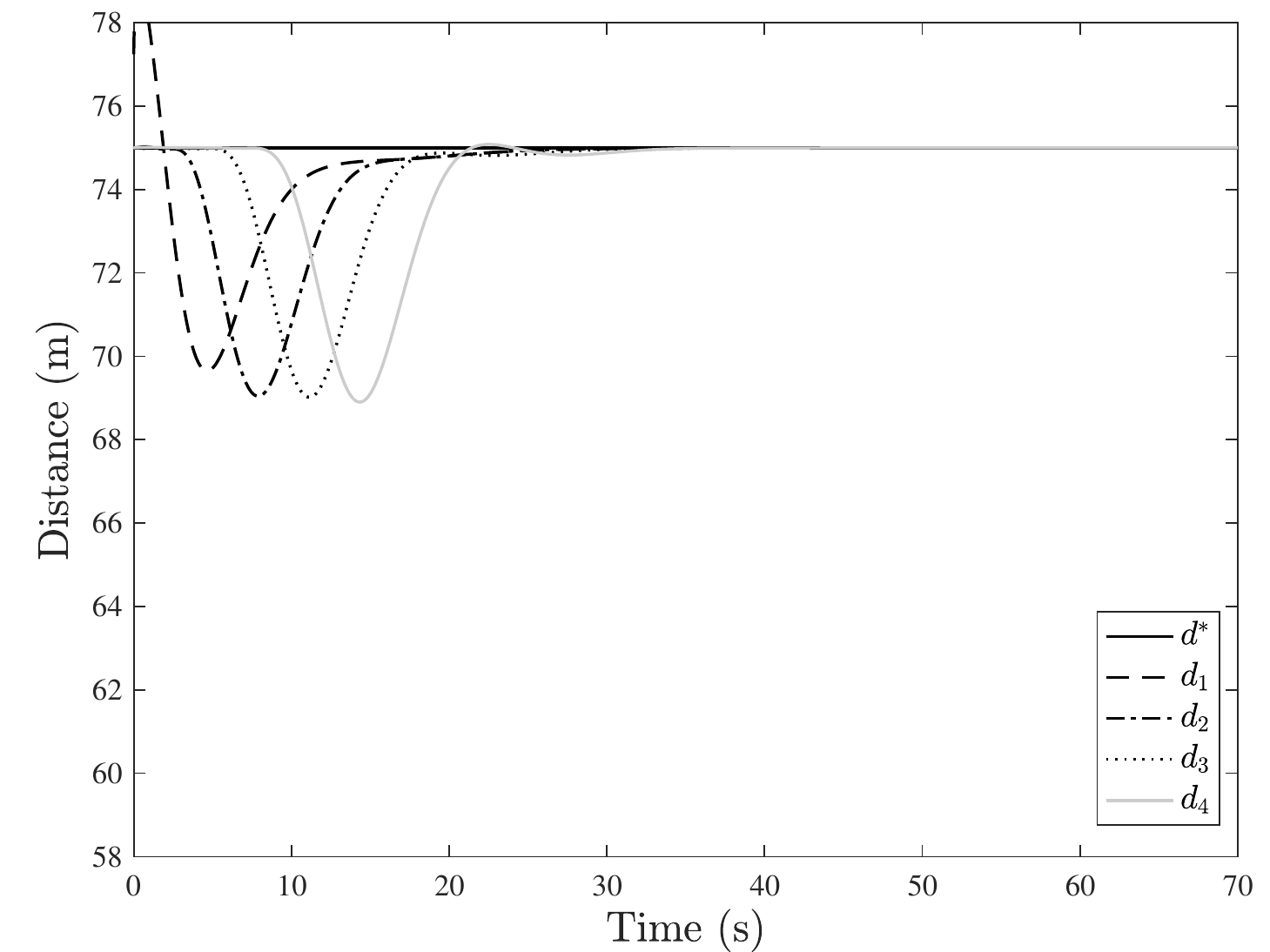}}
	\subfigure[Regular Method Path Errors]{\includegraphics[width = 0.48\textwidth]{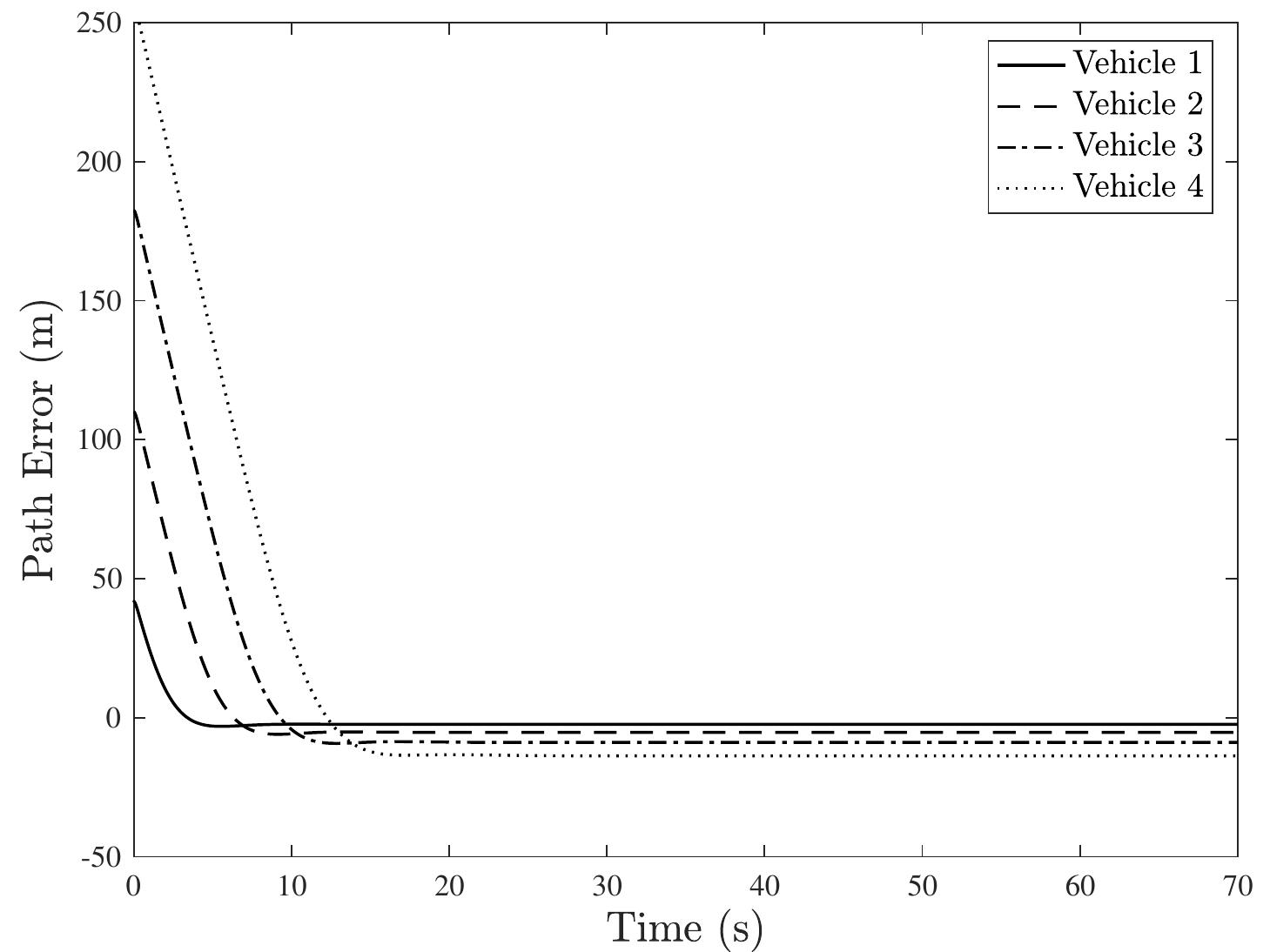}}
	\subfigure[Sine Method Path Errors]{\includegraphics[width = 0.48\textwidth]{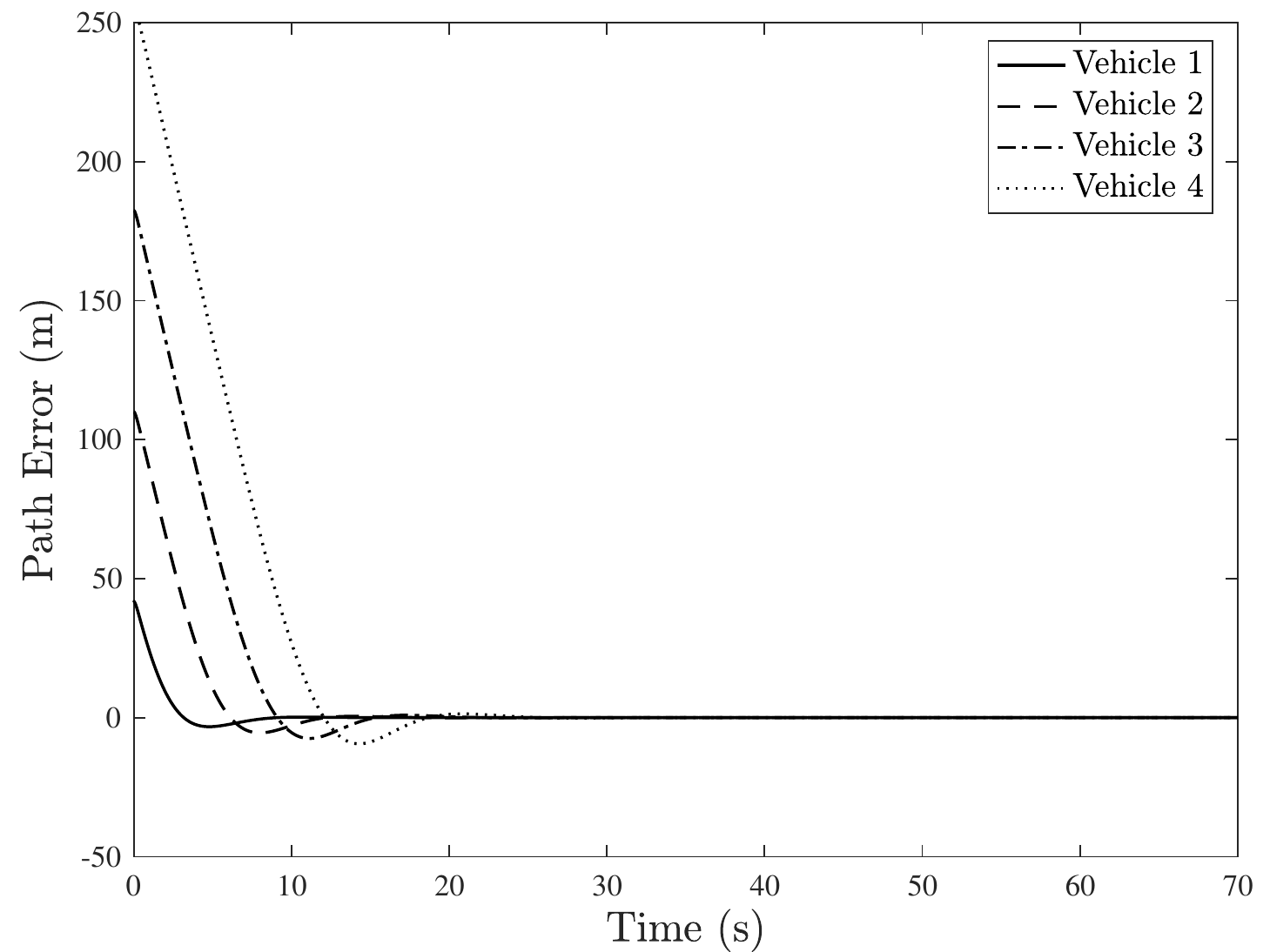}}
	\caption{Platooning Simulation Results.}
	\label{fig:plat_comp_sim}
\end{figure}

For platooning, convergence was shown analytically to occur for the sine method of trajectory shaping.  The regular method has been shown to converge to an offset and each vehicle will converge to an offset of the circle converged to by the vehicle in front of it.  Therefore, the offset adds with each vehicle in the platoon.  This was tested in simulation as shown in Fig.~\ref{fig:plat_comp_sim} and in the video~\url{https://youtu.be/GLPrj4l3o38}.

\begin{figure}[!h]
	\centering
	\subfigure[Regular Method Velocities]{\includegraphics[width = 0.48\textwidth]{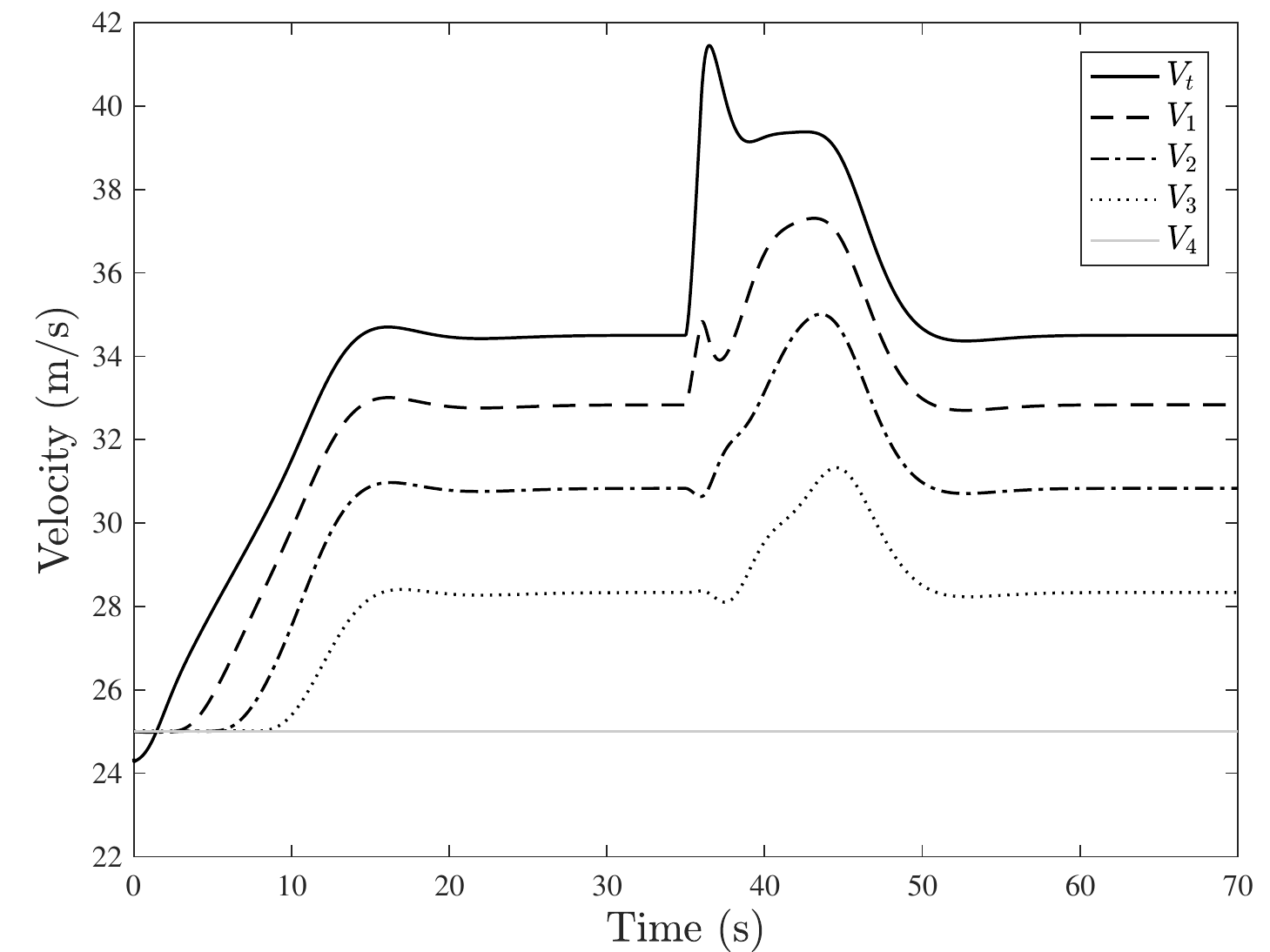}}
	\subfigure[Sine Method Velocities]{\includegraphics[width = 0.48\textwidth]{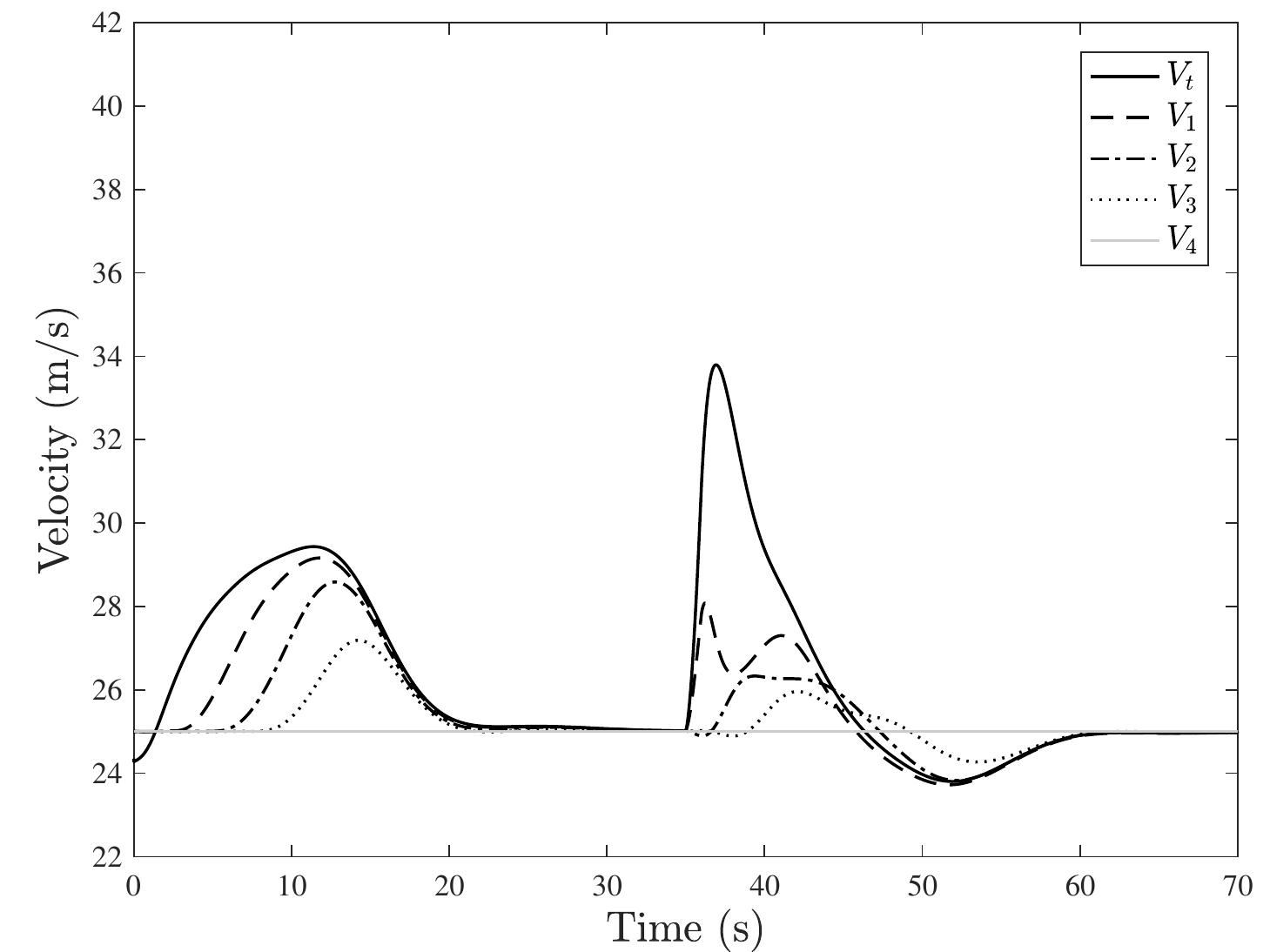}}
	\subfigure[Regular Method Distances]{\includegraphics[width = 0.48\textwidth]{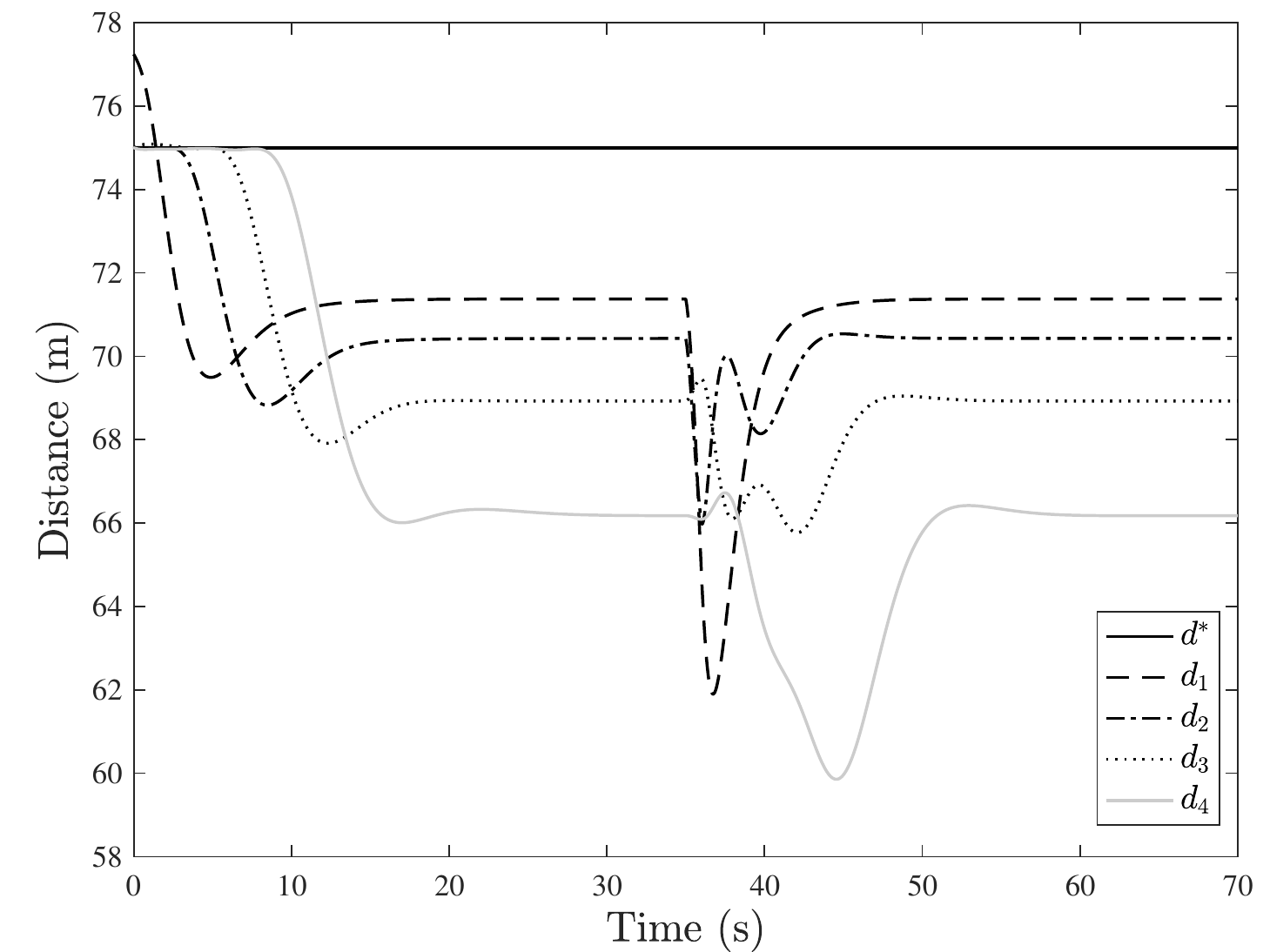}}
	\subfigure[Sine Method Distances]{\includegraphics[width = 0.48\textwidth]{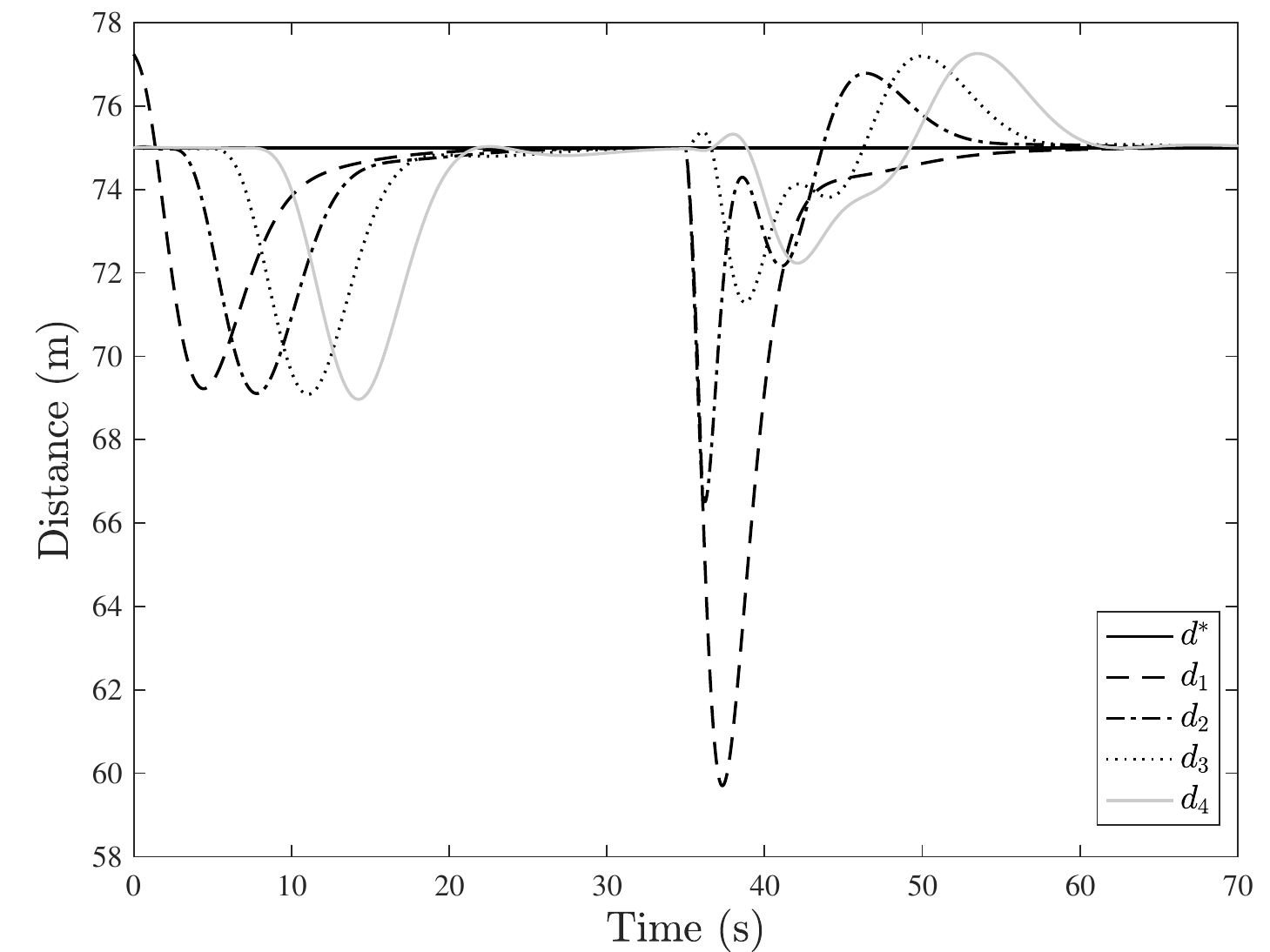}}
	\subfigure[Regular Method Path Errors]{\includegraphics[width = 0.48\textwidth]{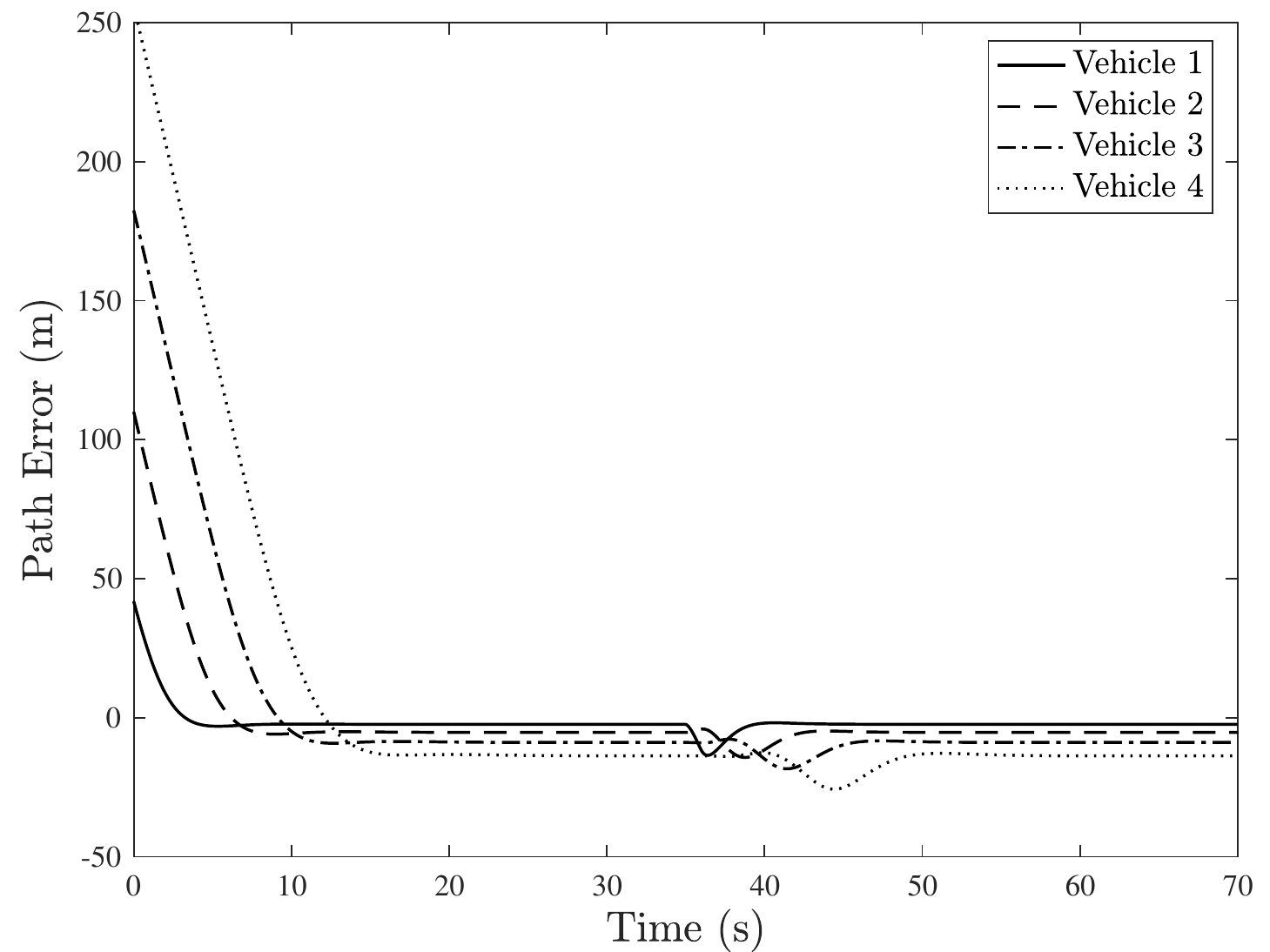}\label{fig:reg_plat_disturb}}
	\subfigure[Sine Method Path Errors]{\includegraphics[width = 0.48\textwidth]{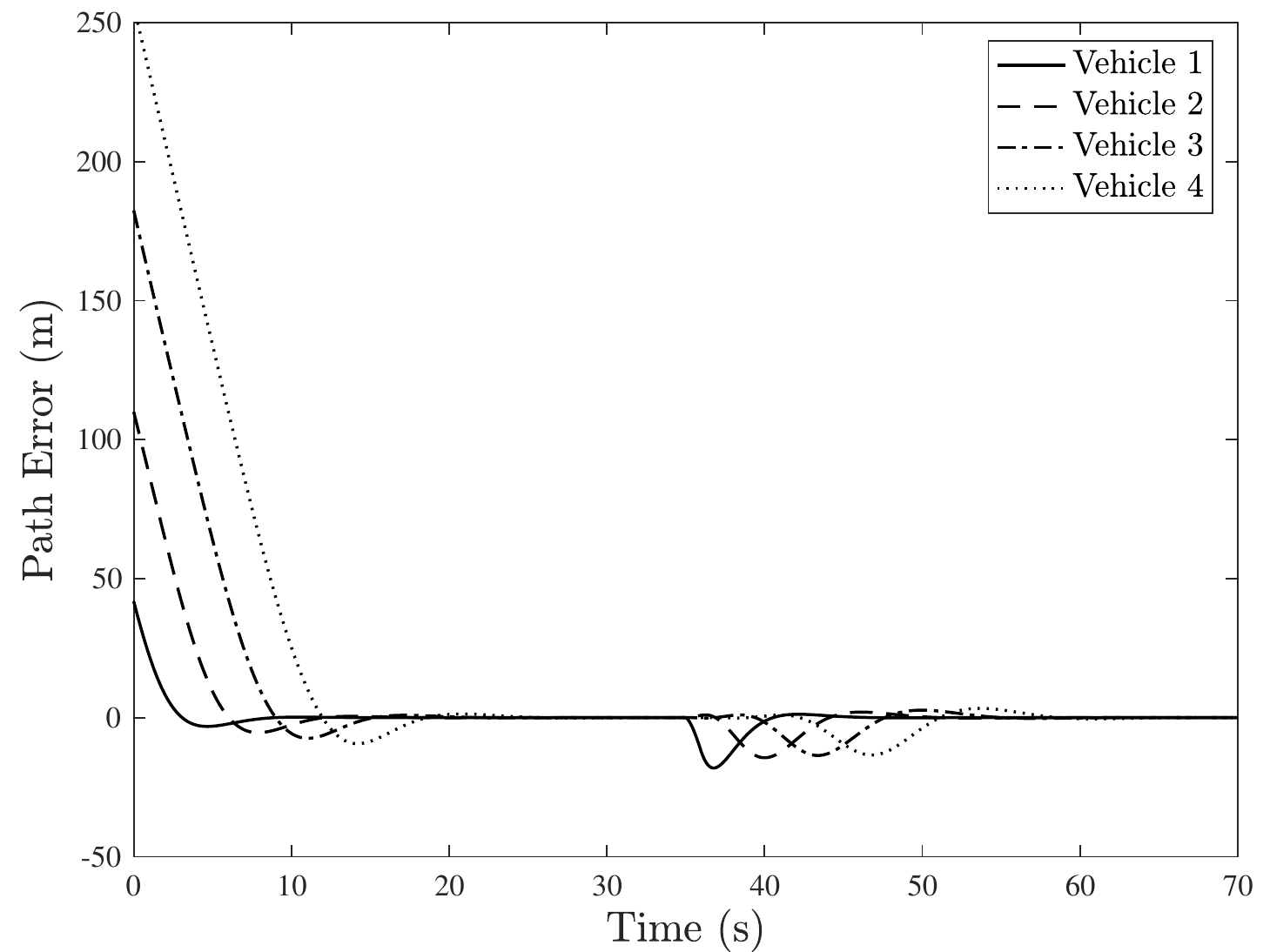}\label{fig:plat_disturb}}
	\caption{Platooning Simulation Results with Front Vehicle Disturbance.}
	\label{fig:plat_disturb_full}
\end{figure}

\begin{figure}[!h]
	\centering
	\subfigure[Regular Method Velocities]{\includegraphics[width = 0.48\textwidth]{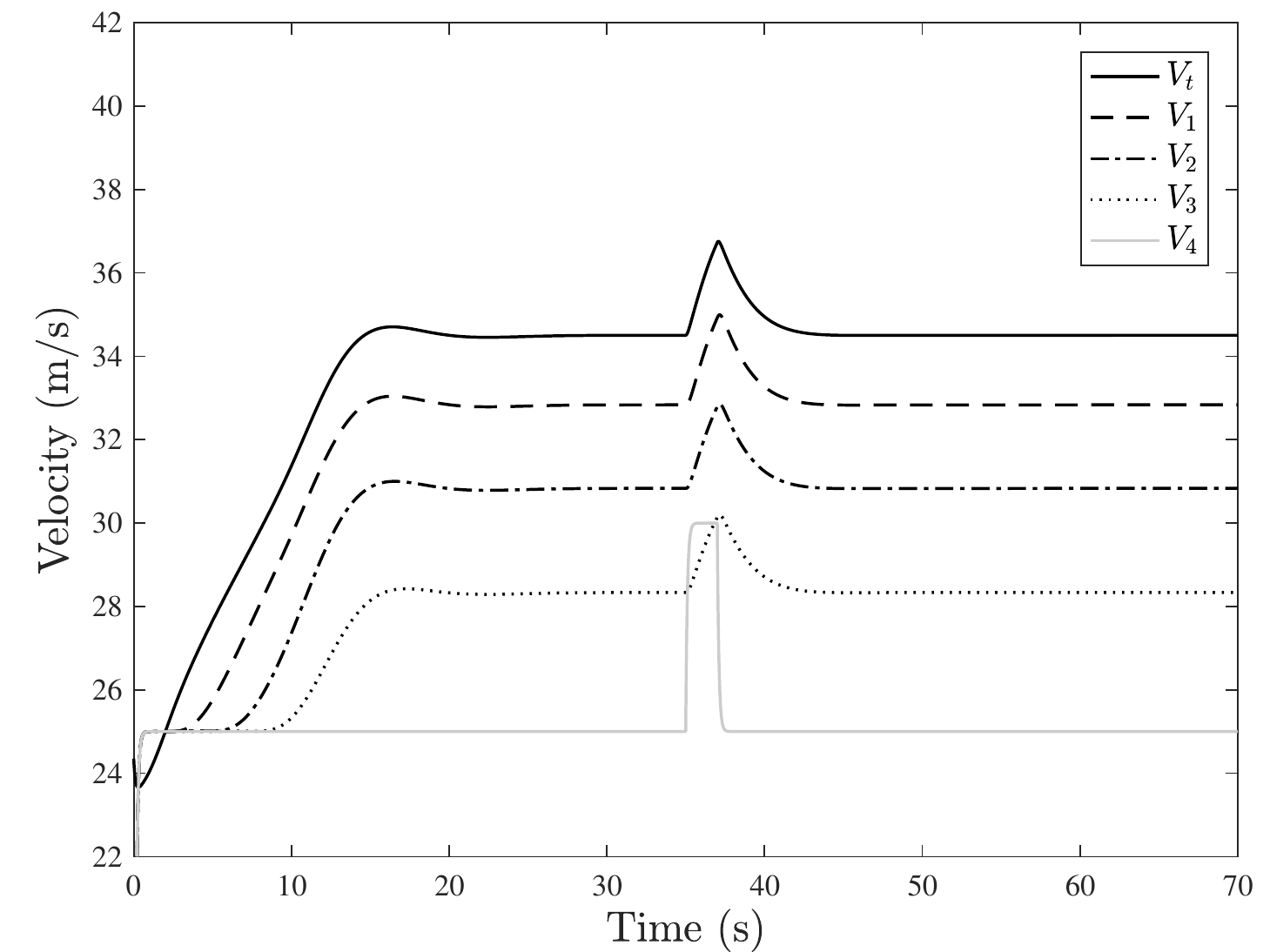}}
	\subfigure[Sine Method Velocities]{\includegraphics[width = 0.48\textwidth]{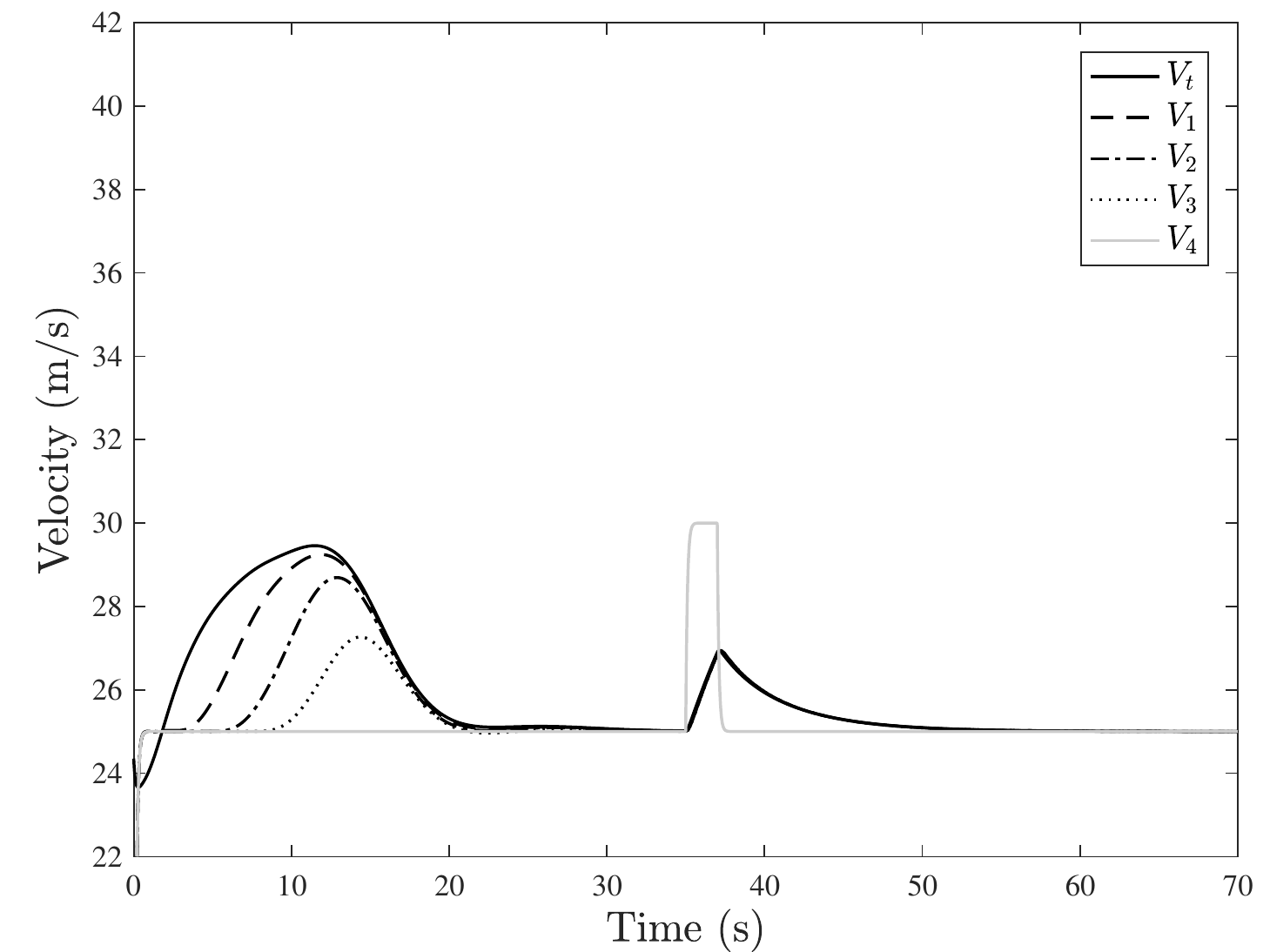}}
	\subfigure[Regular Method Distances]{\includegraphics[width = 0.48\textwidth]{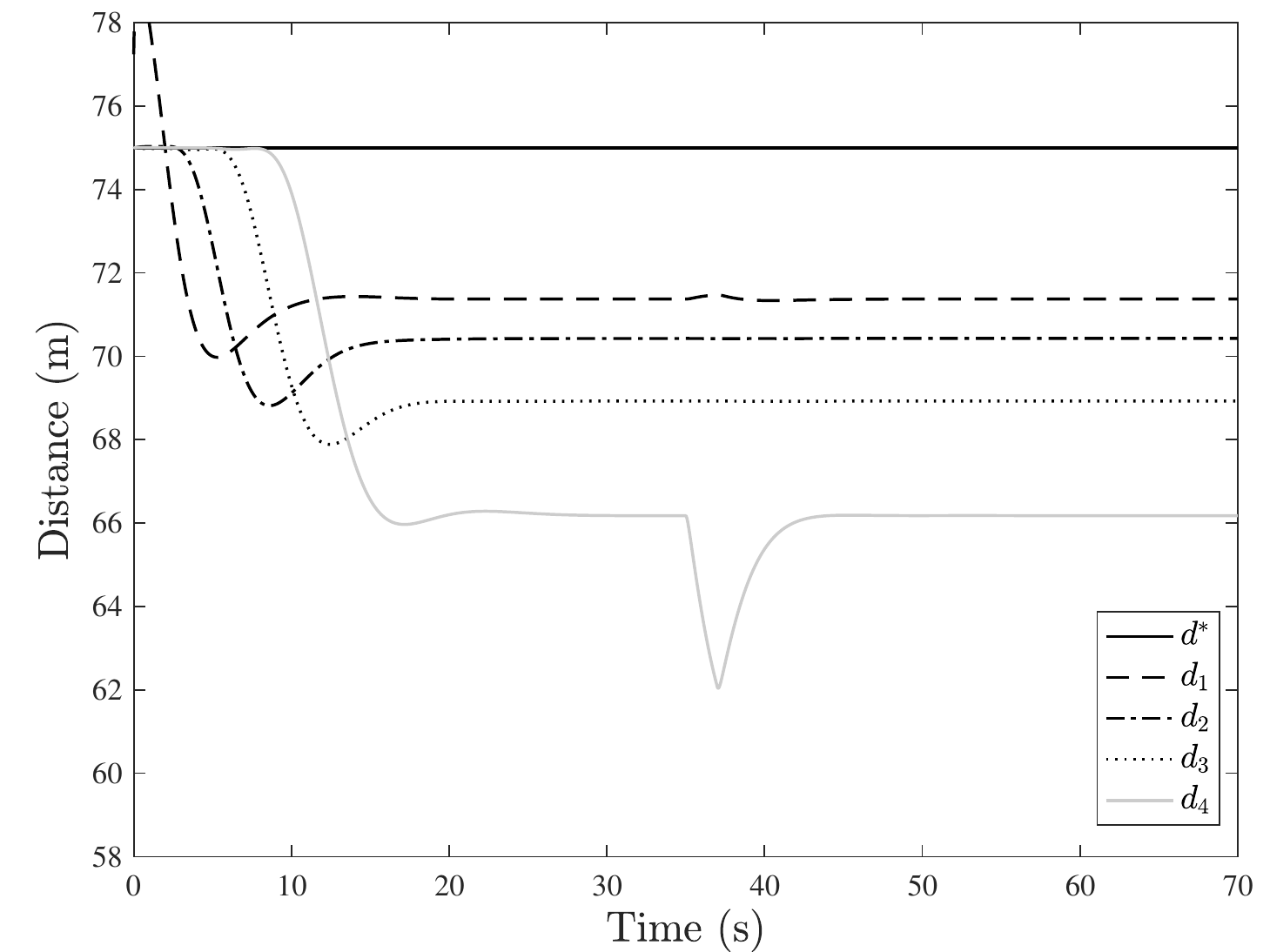}}
	\subfigure[Sine Method Distances]{\includegraphics[width = 0.48\textwidth]{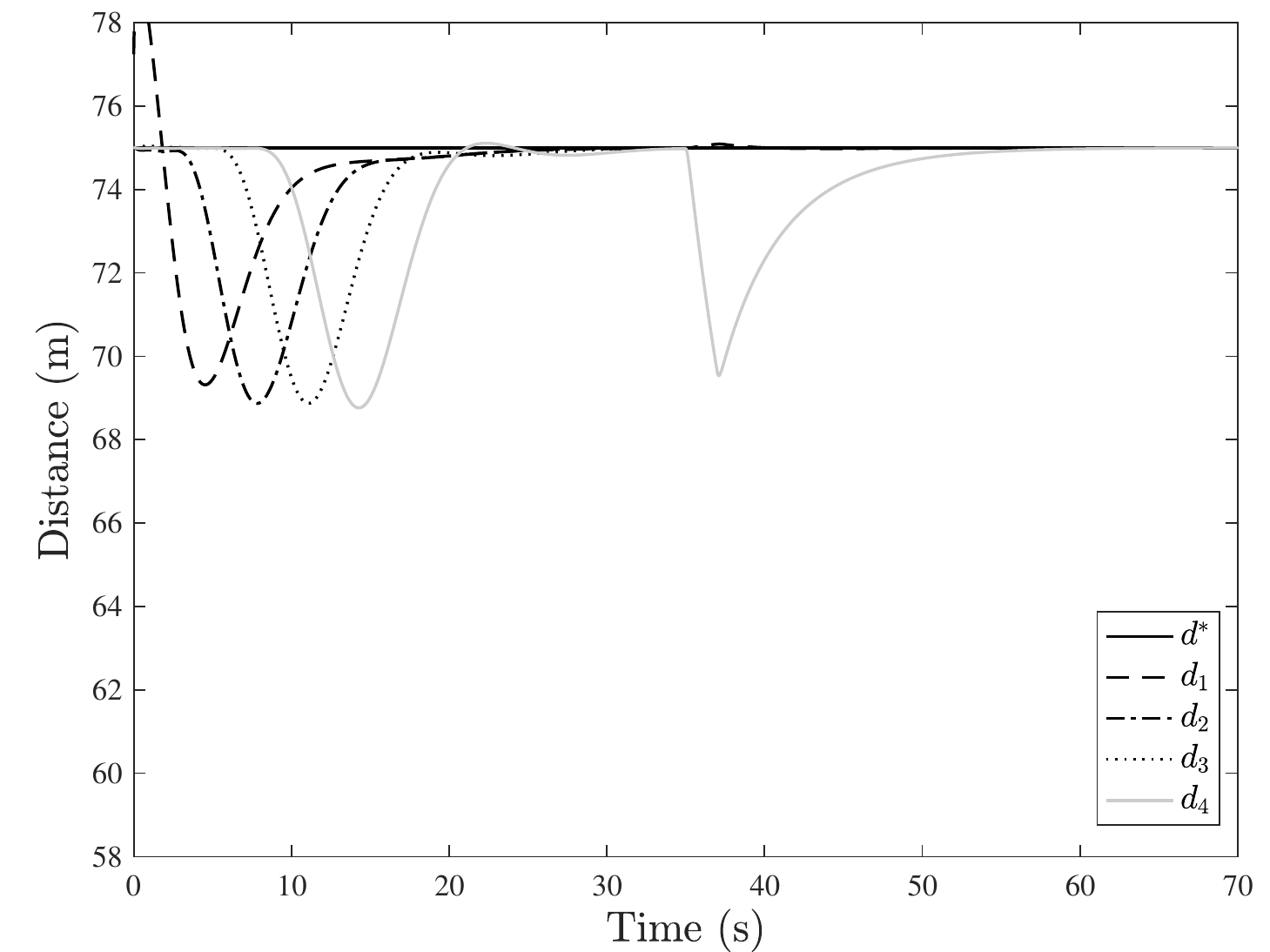}}
	\subfigure[Regular Method Path Errors]{\includegraphics[width = 0.48\textwidth]{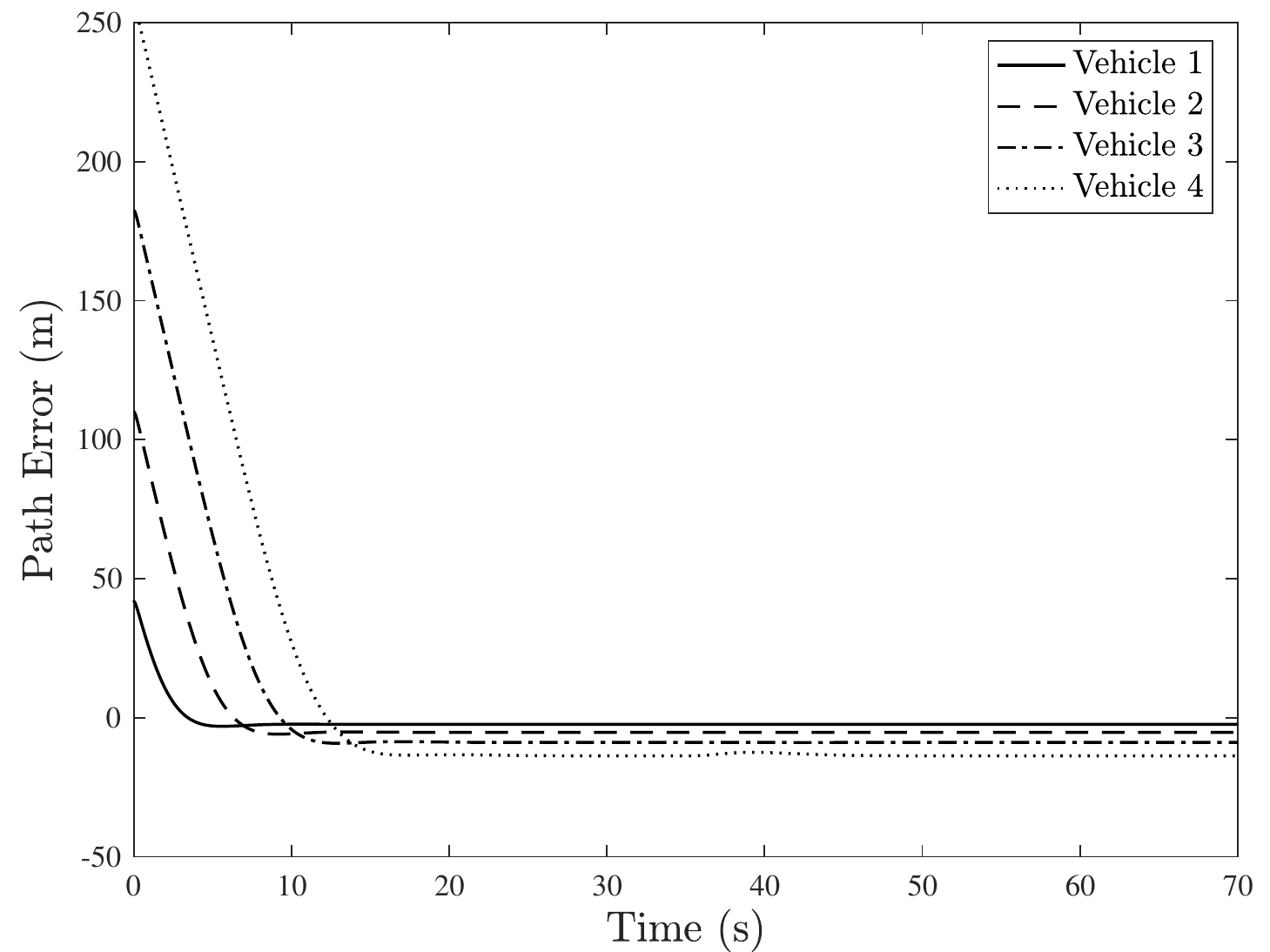}\label{fig:reg_plat_disturb_vel}}
	\subfigure[Sine Method Path Errors]{\includegraphics[width = 0.48\textwidth]{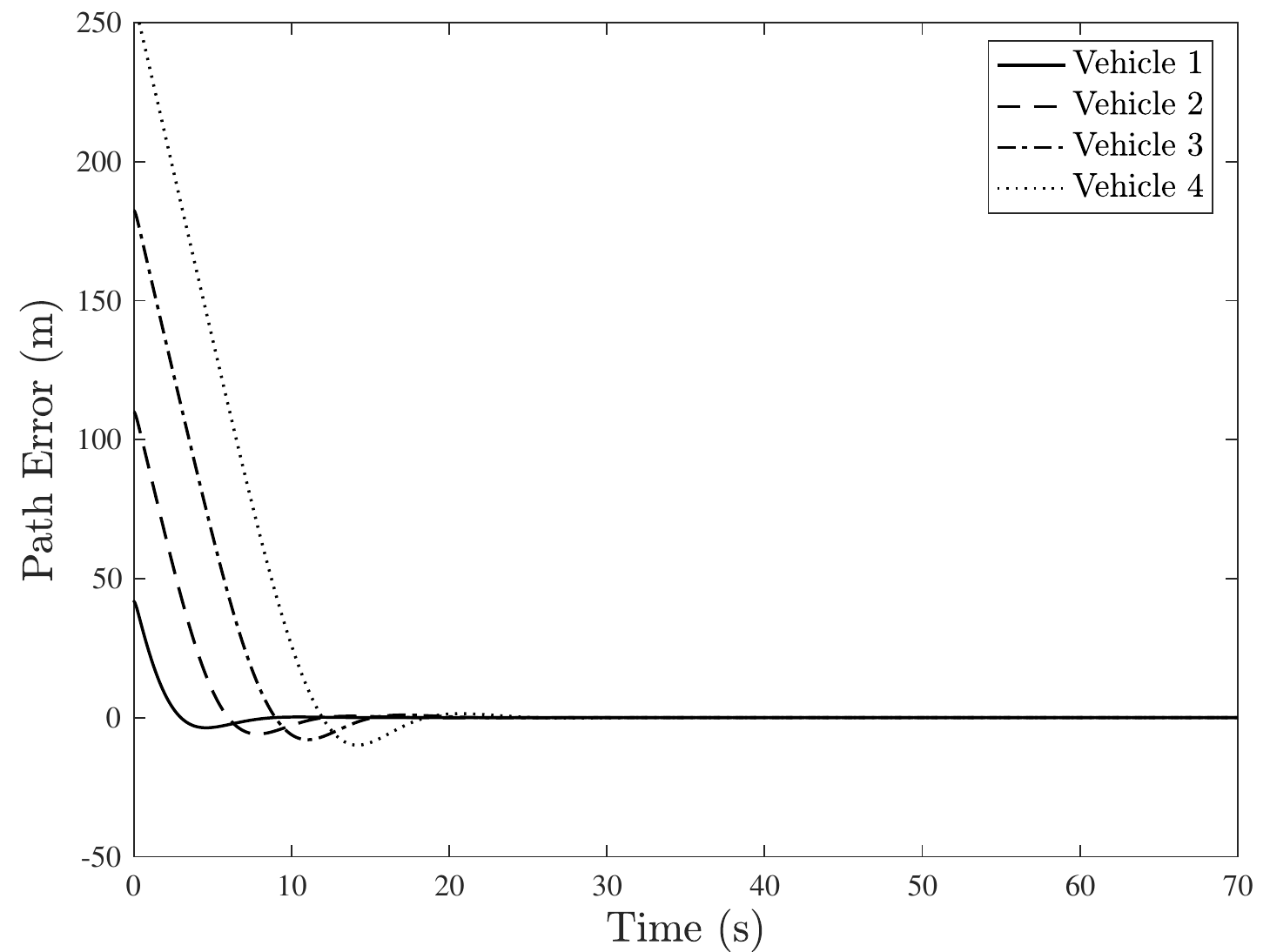}\label{fig:plat_disturb_vel}}
	\caption{Platooning Simulation Results with Back Vehicle Velocity Disturbance}
	\label{fig:plat_veldisturbact_full}
\end{figure}

As can be seen, the sine method had every vehicle converge to the desired path at the desired spacing.  The regular method converged to a slight offset that can be seen in the velocity and distance error plots, but is more difficult to see on the path error plot.  The last vehicle in the regular method converged to a circle with an offset  of $11.35~\textrm{m}$.  This would equate to vehicle 4 being several lanes over on the highway.  
(an important property for platooning) 
It was also desired to see how the system reacts to a disturbance.  The acceleration disturbance was created by adding a extremely large lateral acceleration of $35~\textrm{m/s}^2$ to a vehicle for $1$ second when the simulation reaches $35$ seconds.  The velocity disturbance was created by add $5~\textrm{m/s}$ of velocity to a vehicle for $1$ second when the simulation reaches $35$ seconds.  The results presented in Fig.~\ref{fig:plat_disturb} and Fig.~\ref{fig:plat_disturb_vel} show that the vehicles again converge back to the path.
It can be seen in Fig~\ref{fig:plat_disturb_full} (c)  that for regular case vehicles converge at $d<d^*$ and therefore the velocity of the vehicles are higher  then desired velocity because $V_{i}=V_{i+1}\frac{d^*}{d_i}$ which leads to smaller turn rate command $\dot \gamma_{v_i}=\frac{a_i}{V_i}$ for same applied lateral acceleration disturbance $a$ therefore the maximum deviation on path errors seems higher for the modified trajectory shaping when provided same disturbance.

Simulation videos for ten vehicles are available at \url{https://sharmarajnikant.wordpress.com/projects-2/autonomous-vehicle-platooning/}.

Keeping the above results in mind, we can conclude that the sine or modified trajectory shaping method has overall better convergence properties  as compared to the regular trajectory shaping method.  These results clearly  show that the modified trajectory shaping is a better candidate for all platooning scenarios.  Note that  $d^*$ is not a tuning parameter in the platooning scenario but is an input and sine method is able to achieve any desired distance between vehicles.  Therefore we only consider the modified or sine method for obtaining the hardware results that will be detailed in the next section.

\subsection{Hardware Results}
Due to space constraints of the testing facilities at Utah State University, 6 small robots were set to run at a desired velocity of 0.35~m/s on a 1~m radius path.  Pololu's m3pi robot was selected for these experiments.  This is a differential steering, two-wheeled vehicle, so the lateral acceleration was mapped to the differential velocity of the right and left side of the robot as shown in \eqref{eq:vr} and \eqref{eq:vl} where $V_r$ is the velocity of the right side of the robot, $V_l$ is the velocity of the left side of the robot, $V_c$ is the commanded velocity, and $W$ is the distance between the two wheels.  
\begin{align}
	\label{eq:vr}
	V_r& = V_c \left( 1 + \frac{W}{2 R_t} \right)\\
	\label{eq:vl}
    V_l &= V_c \left( 1 - \frac{W}{2 R_t} \right)
\end{align}

\begin{figure}[!h]
    \centering
    \subfigure[Pololu m3pi Robot with Custom Reflector Template \label{fig:pi3}]{\includegraphics[width = 0.48\columnwidth]{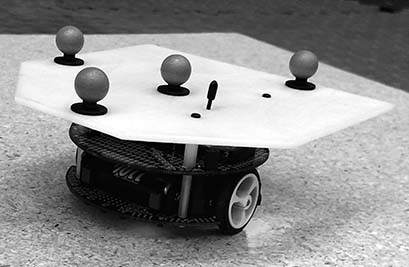}}
    \subfigure[USU's RISC MAAP System\label{fig:maap} ]{\includegraphics[width = 0.48\columnwidth]{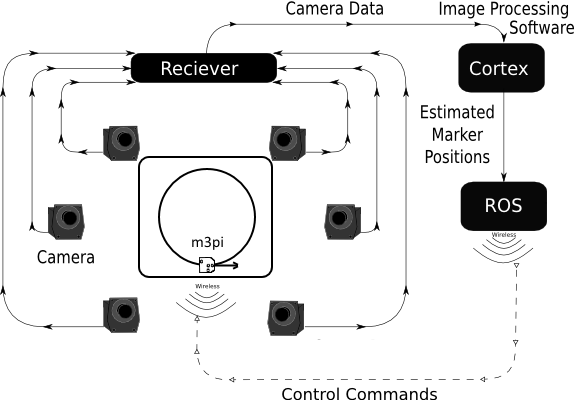}}
	\caption{Hardware System}
\end{figure}

Using Robot Operating System (ROS), and Utah State University's (USU) Robust Intelligent Sensing and Control Multi-Agent Analysis Platform (RISC MAAP) \cite{Maughan2016} the m3pi robot was able to receive position data from a motion capture system, which required each robot to have a unique template of reflective markers as seen in Fig.~\ref{fig:pi3}.  This information was then used to apply the trajectory shaping tests.  This system is able to give position data within around 2mm accuracy and heading data within about 2 degrees at about 200 Hz.  Fig.~\ref{fig:maap} shows the general layout of how the RISC MAAP System works.  The wireless communication was achieved using XBee Series 1 RF Modules.  The ROS framework provides for easy passing of state information from the motion capture system to the main computer with ROS.  The state information could then be translated into motor commands based on the control algorithms and then sent wirelessly to the m3pi robots.

\begin{figure}[!h]
	\centering
	\subfigure[Vehicle Velocity Errors]{\includegraphics[width = 0.48\textwidth]{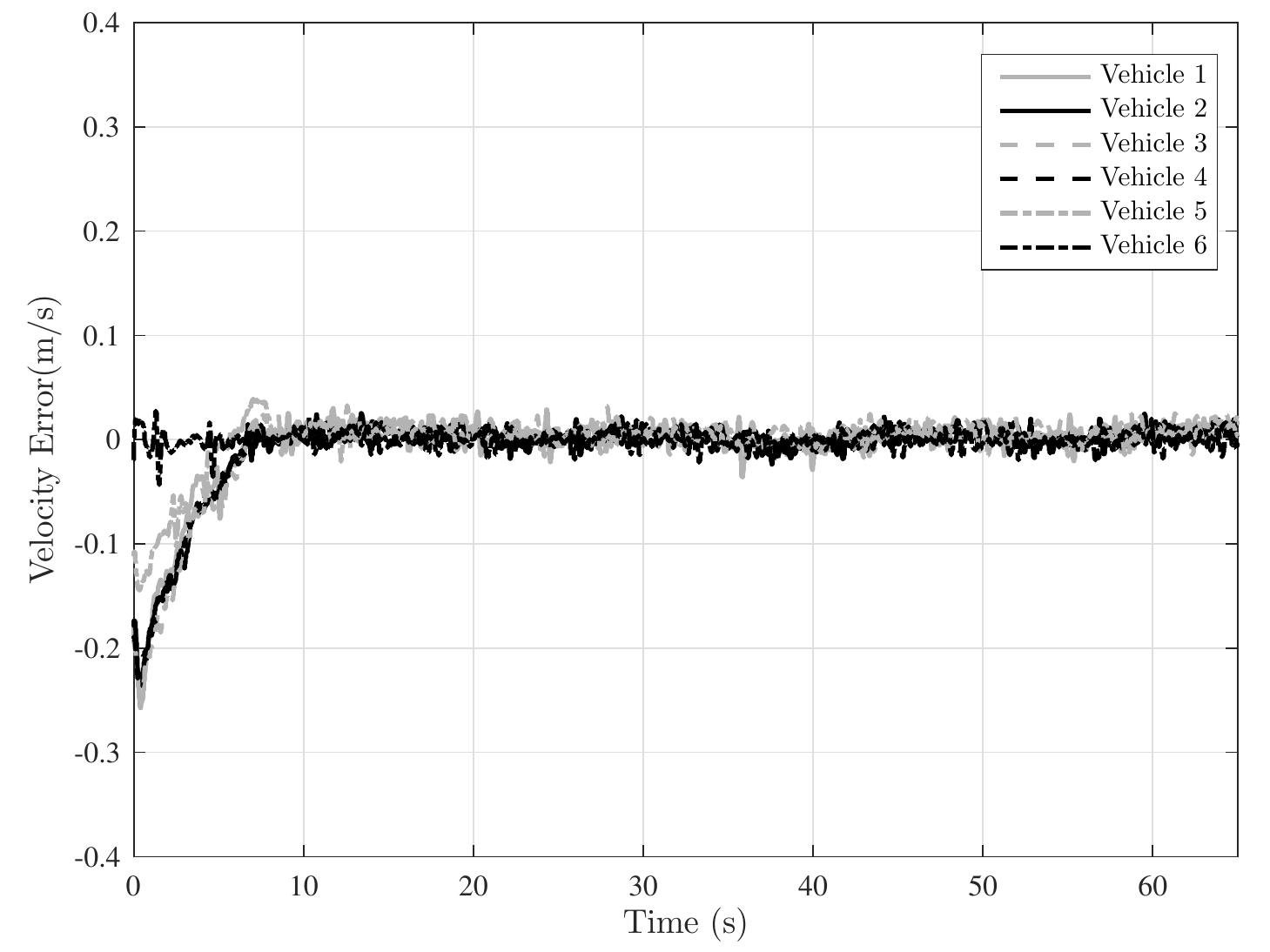}}
	\subfigure[Vehicle Distance Errors]{\includegraphics[width = 0.48\textwidth]{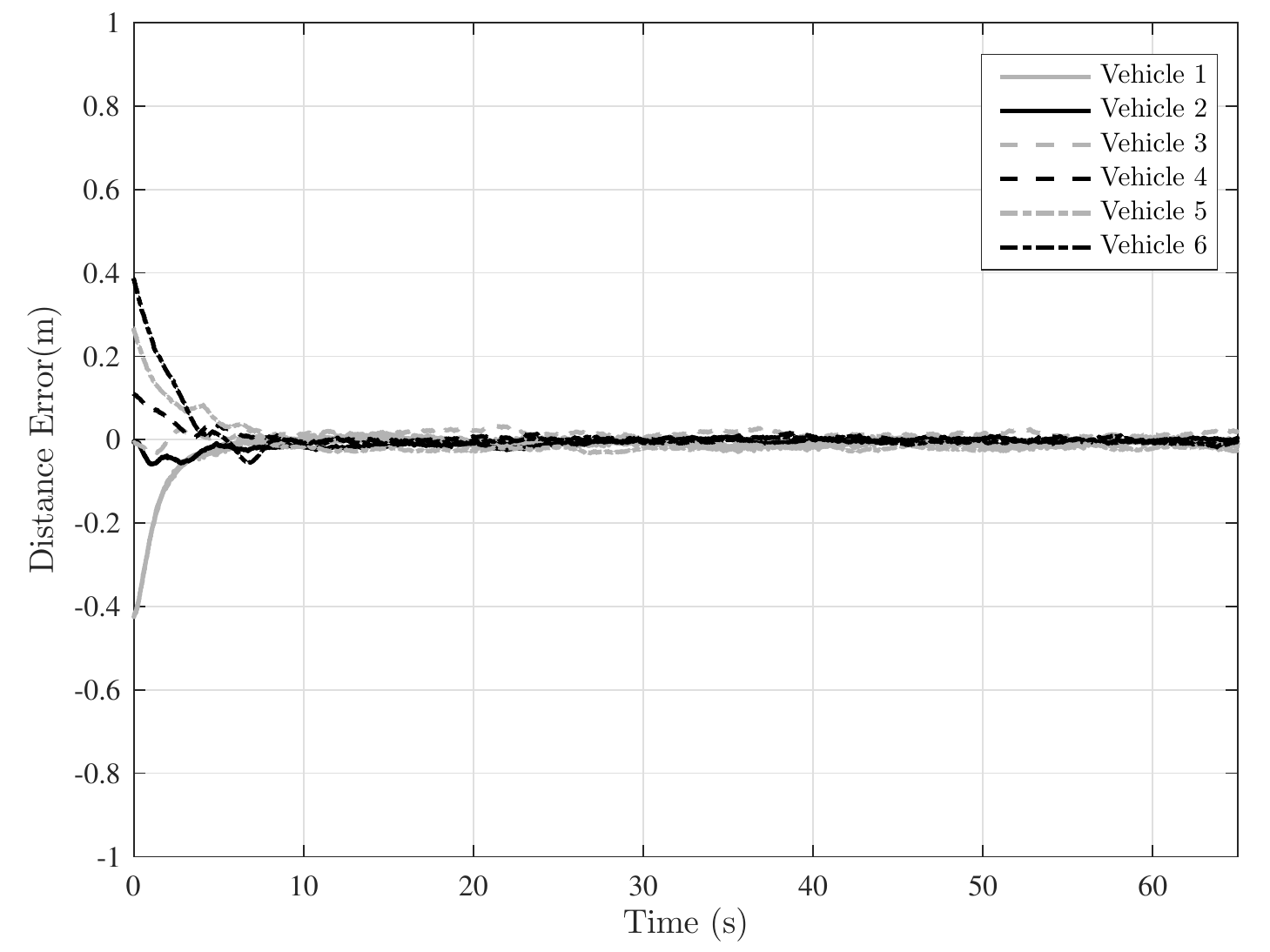}}
	\subfigure[Vehicle to Path Errors]{\includegraphics[width = 0.48\textwidth]{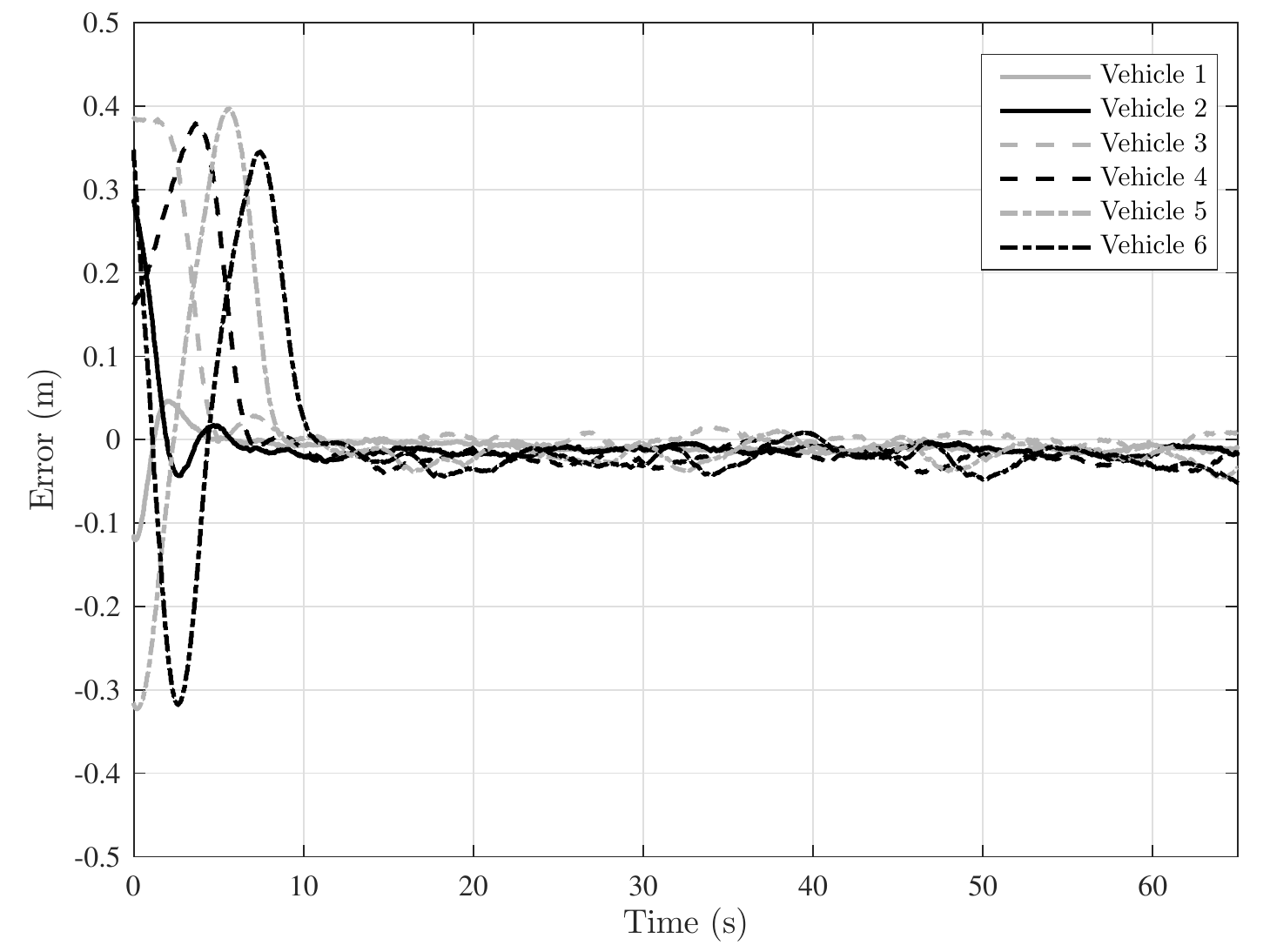}}
	\subfigure[X-Y Positions]{\includegraphics[width = 0.48\textwidth]{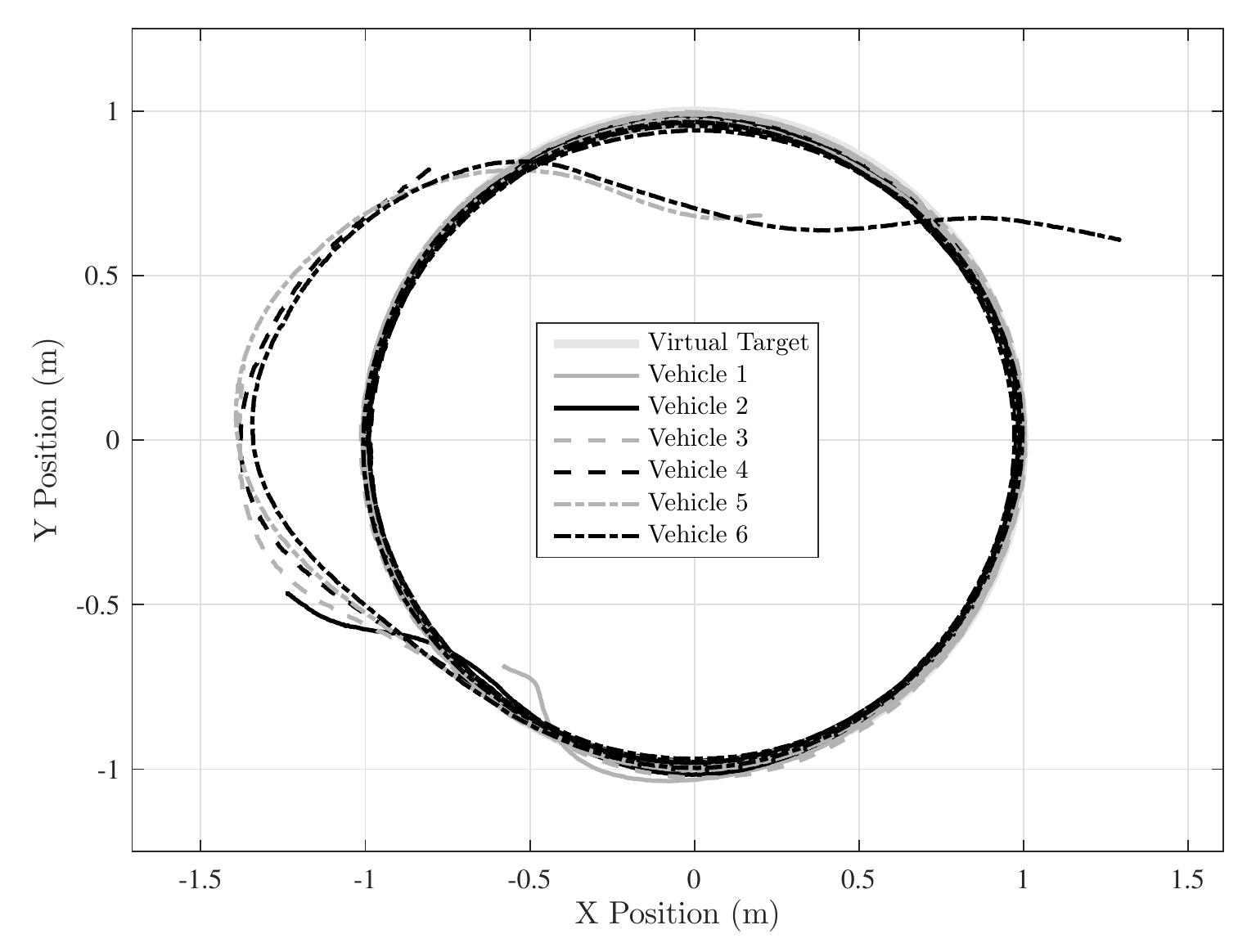}}
	\caption{Platooning Hardware Results: Platoon Behavior at $d^* = 0.7$ m}
	\label{fig:plat_hw}
\end{figure}

\begin{figure}[!h]
	\centering
	\subfigure[Vehicle Velocity Errors]{\includegraphics[width = 0.48\textwidth]{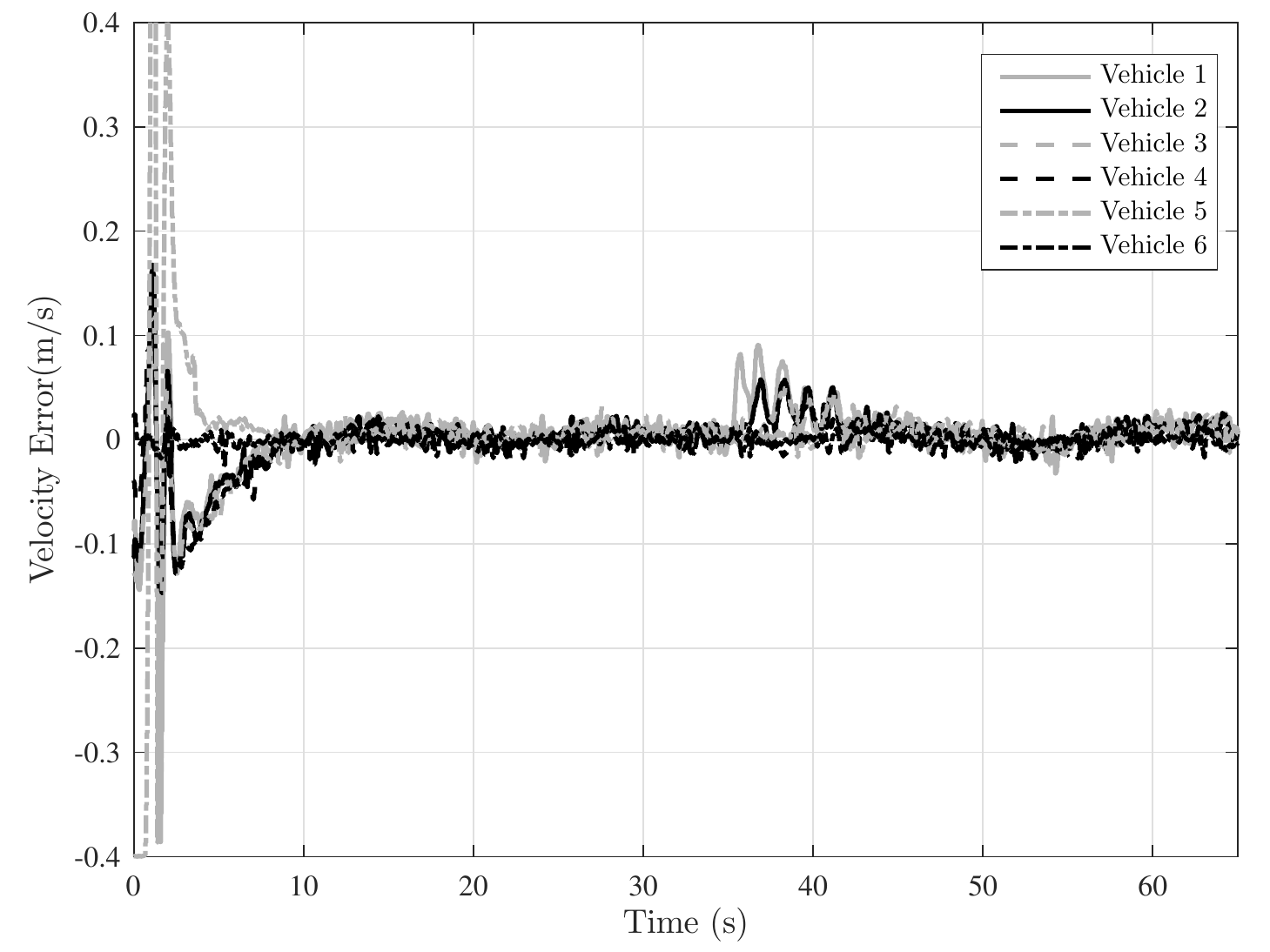}}
	\subfigure[Vehicle Distance Errors]{\includegraphics[width = 0.48\textwidth]{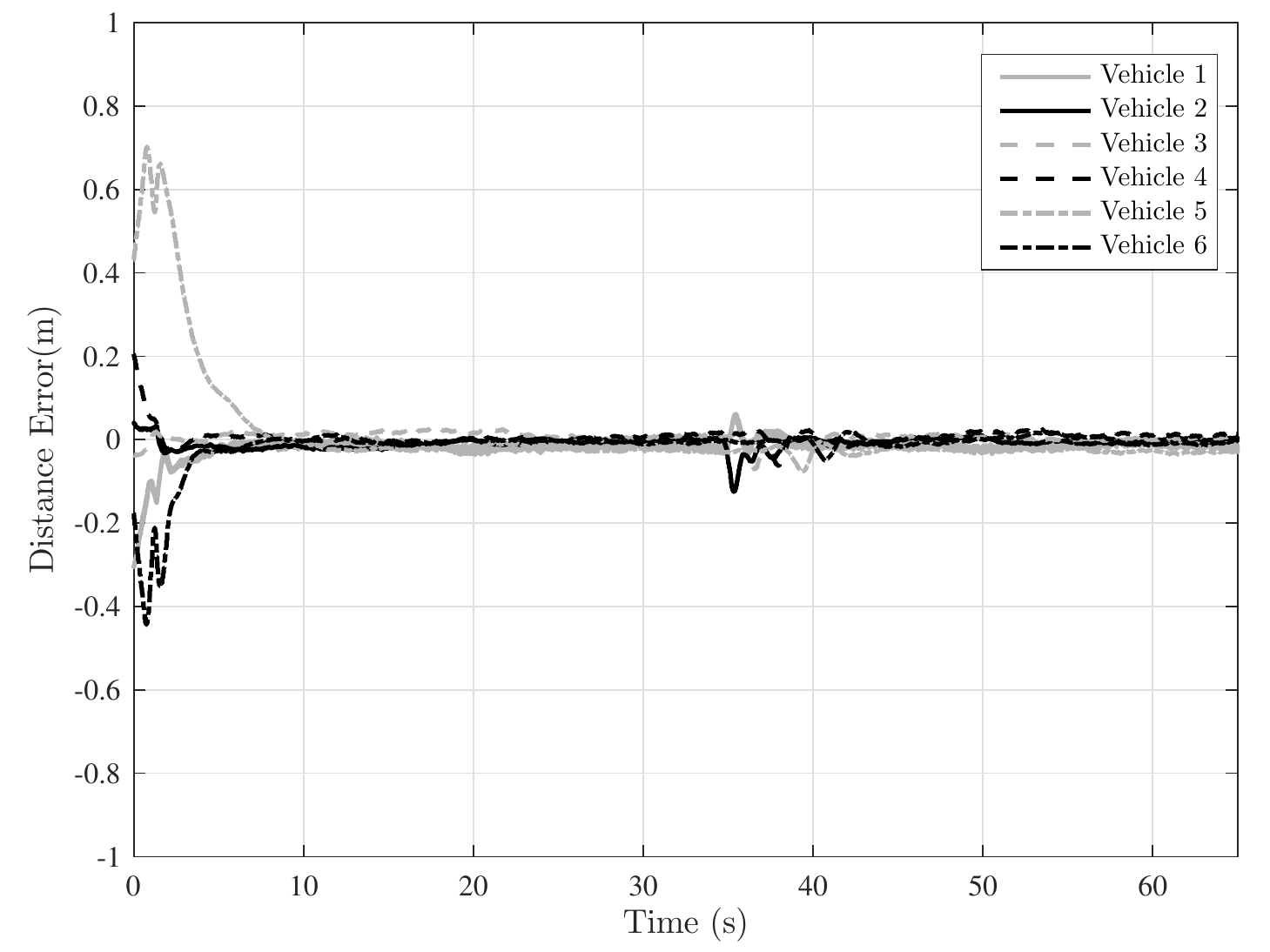}}
	\subfigure[Vehicle to Path Errors]{\includegraphics[width = 0.48\textwidth]{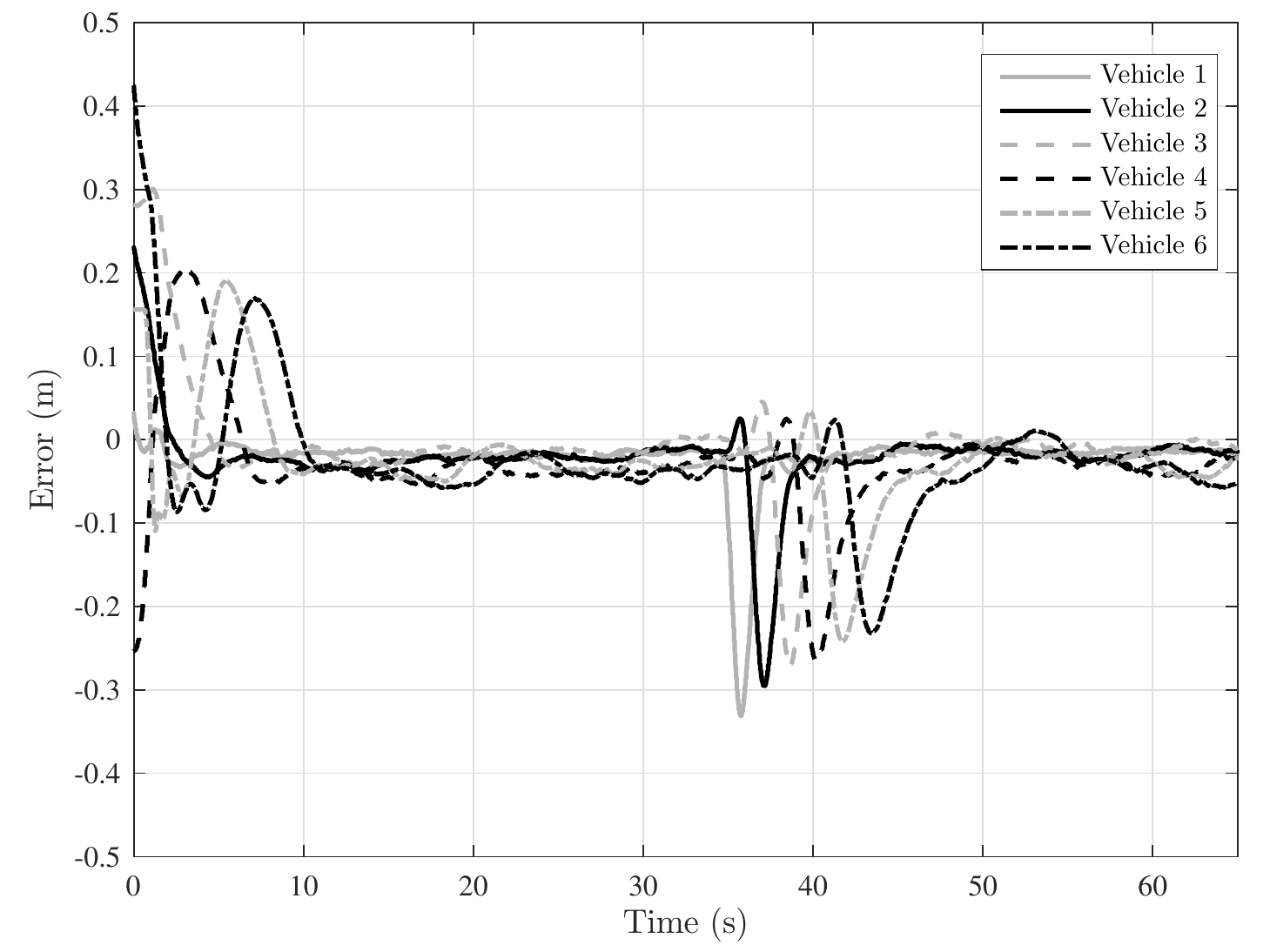}}
	\subfigure[X-Y Positions]{\includegraphics[width = 0.48\textwidth]{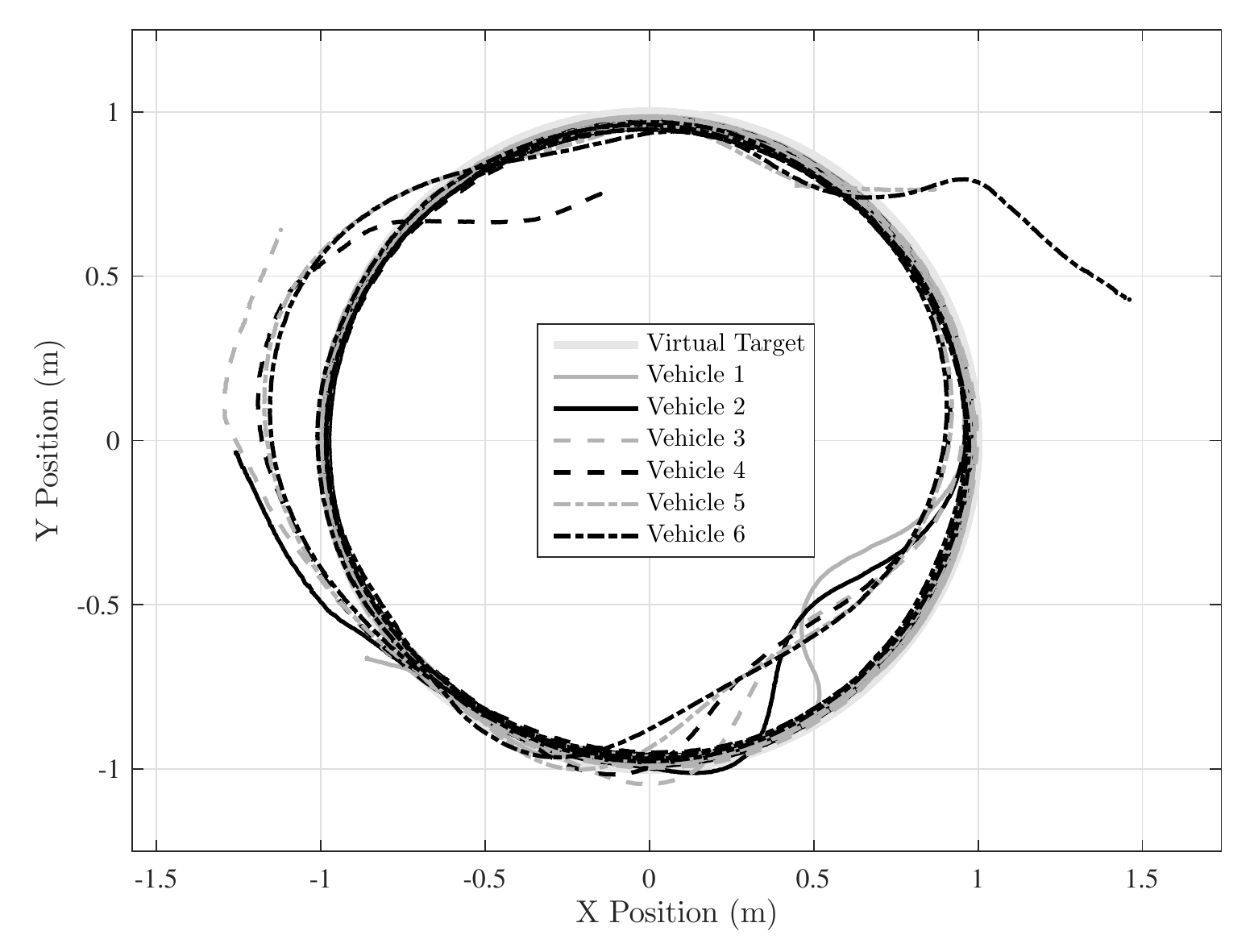}}
	\caption{Platooning Hardware Results: Platoon Behavior with Front Vehicle Disturbance at $d^* = 0.7$ m}
	\label{fig:plat_hw_dist}
\end{figure}

\begin{figure}[!h]
	\centering
	\subfigure[Vehicles Have Converged to Path]{\includegraphics[width = 0.32\textwidth]{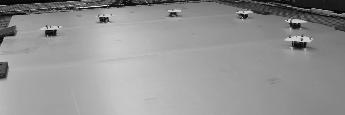}}
	\subfigure[Vehicle 1 Disturbance at 30s]{\includegraphics[width = 0.32\textwidth]{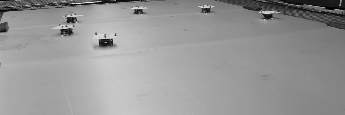}}
	\subfigure[Convergence After Disturbance]{\includegraphics[width = 0.32\textwidth]{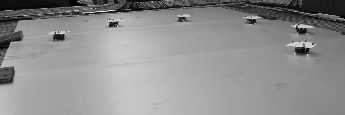}}
	\caption{Hardware Test Images: Disturbance to Front Vehicle. See full experimental video at \url{https://youtu.be/GLPrj4l3o38}.}
	\label{fig:dist}
\end{figure}

\begin{figure}[!h]
	\centering
	\subfigure[Vehicle Velocity Errors]{\includegraphics[width = 0.48\textwidth]{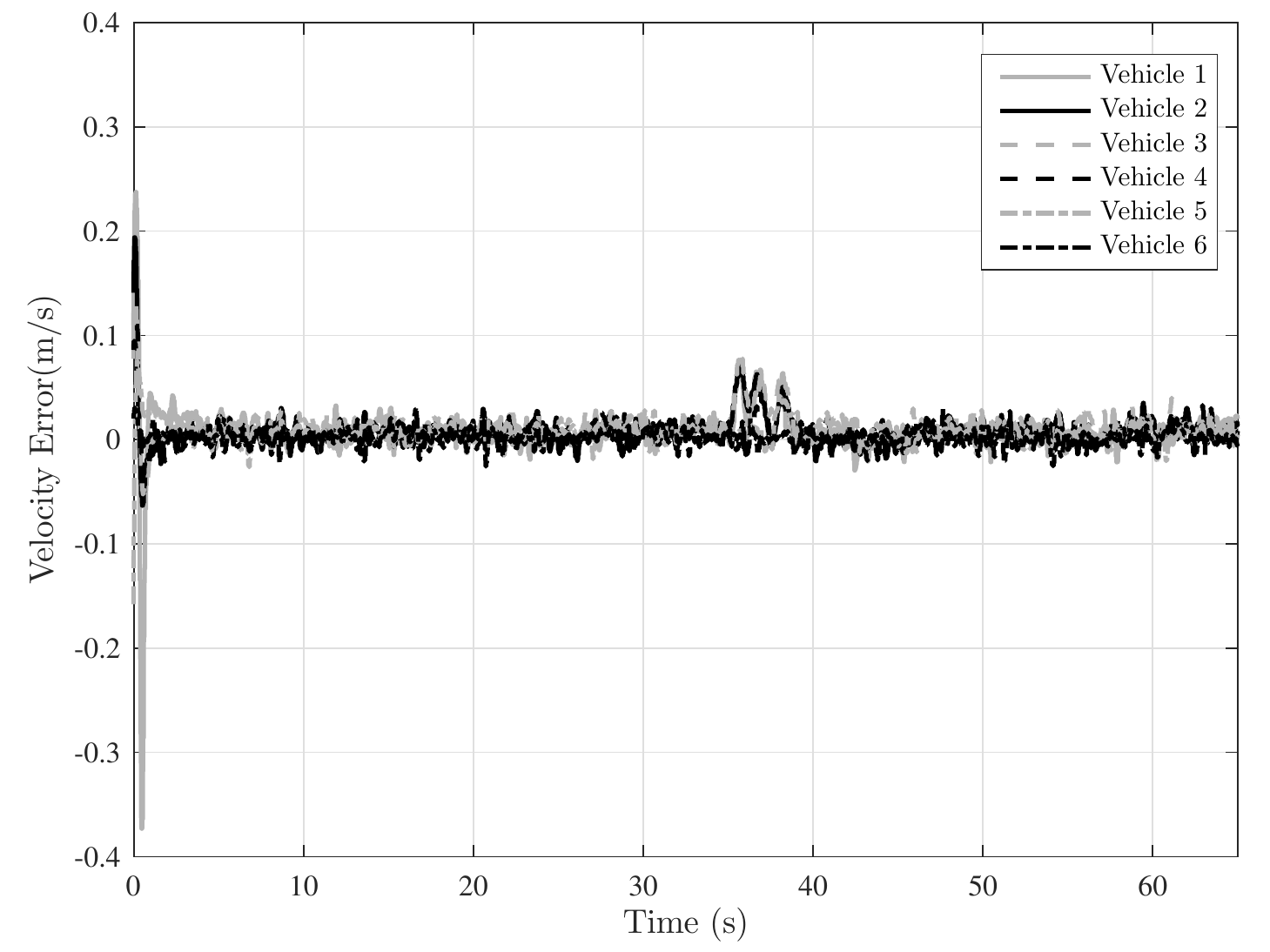}}
	\subfigure[Vehicle Distance Errors]{\includegraphics[width = 0.48\textwidth]{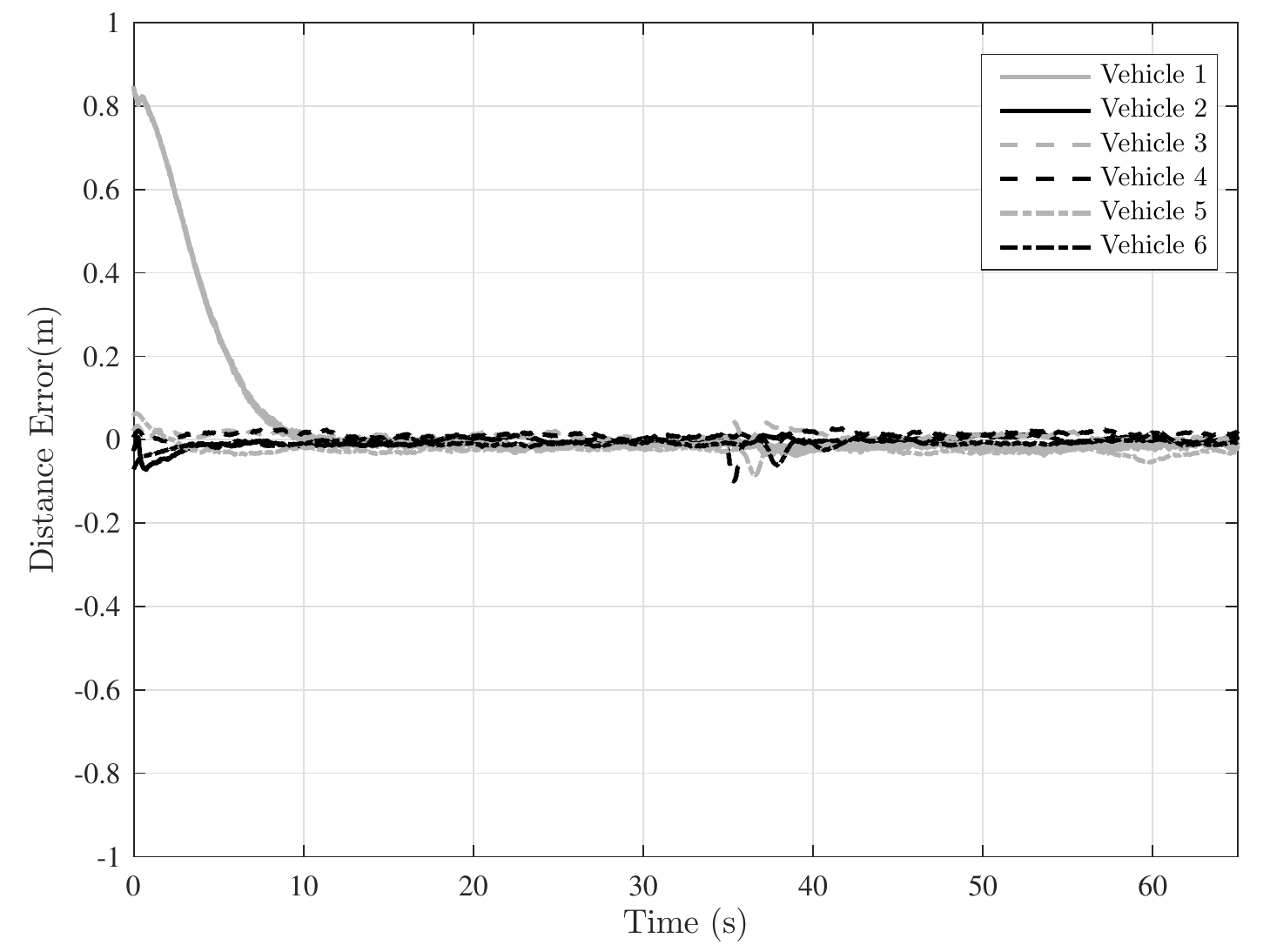}}
	\subfigure[Vehicle to Path Errors]{\includegraphics[width = 0.48\textwidth]{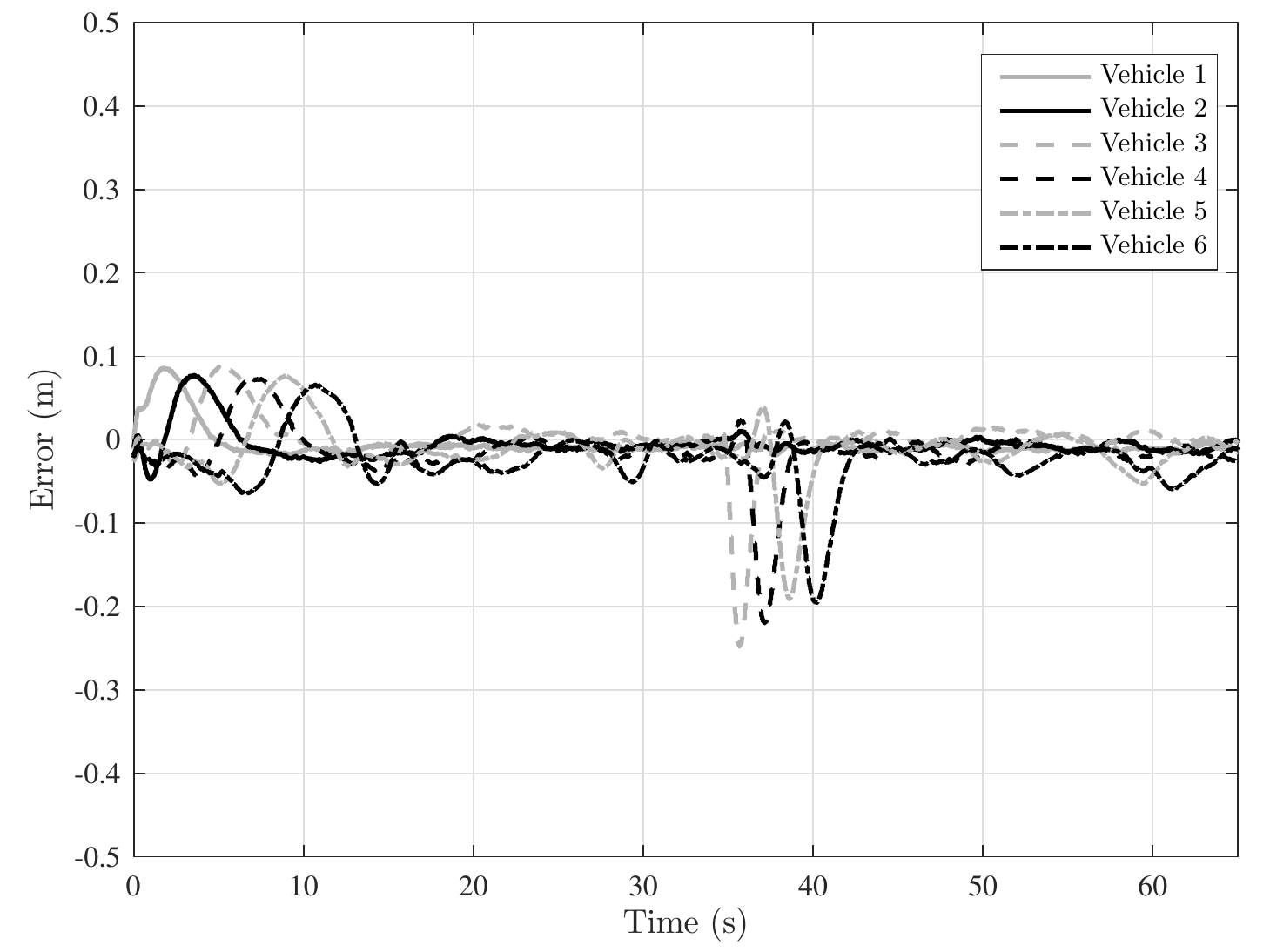}}
	\subfigure[X-Y Positions]{\includegraphics[width = 0.48\textwidth]{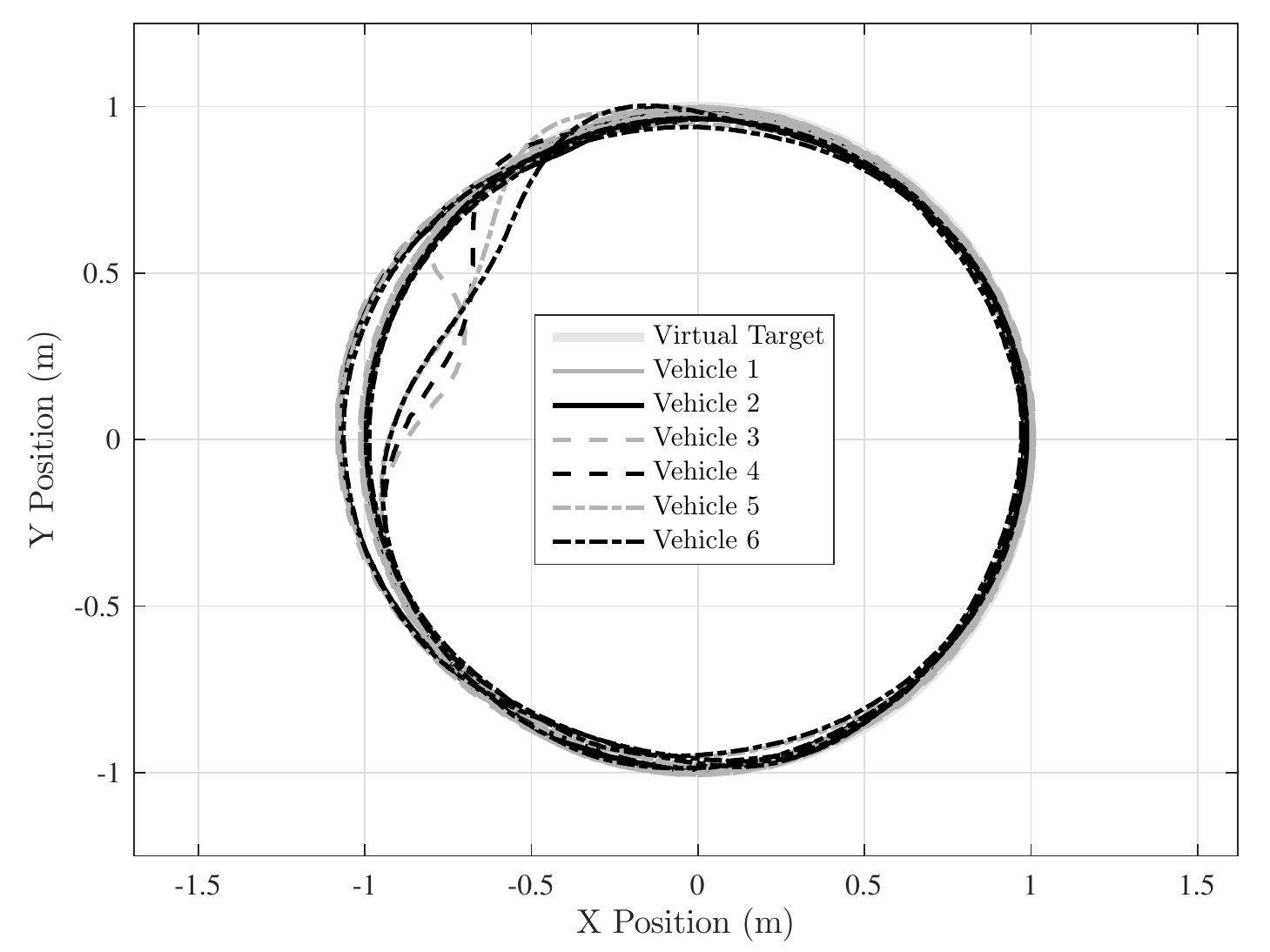}}
	\caption{Platooning Hardware Results: Platoon Behavior with Middle Vehicle Disturbance at $d^* = 0.7$ m}
	\label{fig:plat_hw_dist_mid}
\end{figure}

\begin{figure}[!h]
	\centering
	\subfigure[Vehicles Have Converged to Path]{\includegraphics[width = 0.32\textwidth]{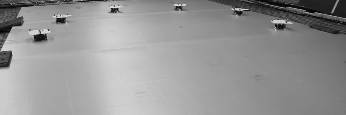}}
	\subfigure[Vehicle 3 Disturbance at 30s]{\includegraphics[width = 0.32\textwidth]{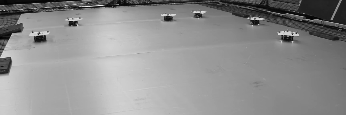}}
	\subfigure[Convergence After Disturbance]{\includegraphics[width = 0.32\textwidth]{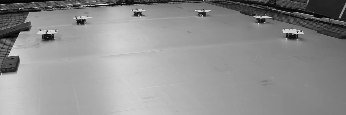}}
	\caption{Hardware Test Images: Disturbance to Middle Vehicle. See full experimental video at\url{https://youtu.be/GLPrj4l3o38}.}
	\label{fig:dist_mid}
\end{figure}

Using a desired vehicle distance of $d^*=0.7~\textrm{m}$, the hardware results are shown in Fig.~\ref{fig:plat_hw}. All the velocity, separation, and path errors converge close to zero and all the robots converges on the path.
As a next step we validate the robustness to disturbance by introducing a disturbance of $1~\textrm{m}/s^2$ in lateral acceleration.  First the disturbance is injected at the front vehicle. The error plots for  this case are shown in Fig.~\ref{fig:plat_hw_dist} that show that the vehicles  converge to the desired path after the disturbance. 
Snapshots of this experiment are shown in Figure~\ref{fig:dist} and complete experimental video can be found at \url{https://youtu.be/GLPrj4l3o38}.
In the second case disturbance is injected in the middle of the platoon in Vehicle 3.  The results shown in Fig.~\ref{fig:plat_hw_dist_mid} show that the path errors of Vehicle 1 and Vehicle 2 are unaffected by the disturbance.  However, they do slow down to allow the others cars to catch up again.  This is due to the flow of information shown in Fig.~\ref{fig:veh}.  The lateral acceleration of each vehicle is only a function of the vehicle in front of it, so any lateral disturbance input to the vehicle would not affect its predecessors. 
Snapshots of this experiment are shown in Figure~\ref{fig:dist_mid} and complete experimental video can be found at  \url{https://youtu.be/GLPrj4l3o38}.

\section{Conclusions}\label{sec: conclusions}

In this paper we first show that the regular trajectory shaping guidance law converges at nonzero offset from the path and is dependent on the value of $d^*$. The trajectory shaping guidance law is modified such that the vehicle converges to circular and straight line paths at zero offset and is independent of the $d^*$ value. It is shown that the modified trajectory shaping guidance is a great candidate for multiple vehicle path following/platooning where only the lead vehicle has the knowledge of the path. The modified trajectory shaping guidance algorithm is validated for path following and platooning using simulation and experimental results. 

For future work, we will investigate string stability, not investigated in this paper, to find out conditions for the string stability.

\bibliography{ref}
\bibliographystyle{IEEEtran}

\end{document}